\documentclass[aps, reprint,groupedaddress,longbibliography,superscriptaddress,nofootinbib]{revtex4-2}
\usepackage{graphicx, tabularx, longtable, nicefrac, amsmath,enumerate, supertabular, eurosym}
\usepackage{hyperref,xcolor}
\hypersetup{breaklinks=true, colorlinks = true, citecolor = blue!30!black, linkcolor = blue!30!black}
\usepackage{booktabs}
\newcommand{\env}[1]{\texttt{#1}}
\usepackage[utf8]{inputenc}
\usepackage[T1]{fontenc}
\usepackage{tikz}
\usepackage{pgfplots}
\pgfplotsset{compat=1.8}
\usepgfplotslibrary{polar}
\definecolor{bblack}{HTML}{363033}
\usepackage{xcolor, booktabs}
\definecolor{grun}{rgb}{0.0, 0.5, 0.0}
\definecolor{amber}{rgb}{1, 0.49, 0.0}
\definecolor{alizarin}{rgb}{0.82, 0.1, 0.26}
\definecolor{dblau}{RGB}{21,50,104}
\definecolor{hblau}{RGB}{132,191,234}
\definecolor{chamois}{RGB}{243,226,216}
\definecolor{altweis}{RGB}{246,244,240}
\definecolor{dgrau}{RGB}{0,101,141}
\definecolor{hgrau}{RGB}{87,87,86}
\usetikzlibrary{patterns, shapes,arrows,positioning, shadows,fit,backgrounds}
\usepackage{setspace}
\usepackage{tabu}
\usepackage{multirow}
\usepackage{xcolor}
\usepackage{calc}
\usepackage{amsmath,amsthm,amssymb,euscript} % AMS-LaTeX  
\usepackage{enumerate,graphicx,setspace}
\usepackage{times}
\usepackage{longtable}
\usepackage{url}
\usepackage{enumitem}
\usepackage{comment}
\usepackage{endnotes}
\usepackage{pgf-pie}
\usepackage{diagbox}
\usepackage{afterpage}
\usepackage{placeins}
\usepackage{bbding}
\usepackage{pgfplots}
\pgfplotsset{compat=1.8}
\usepgfplotslibrary{statistics}

\begin{document}

\title{Recitation tasks revamped? Students' perceptions of smartphone-based experimental and programming tasks in introductory mechanics}

\author{Simon Zacharias Lahme}
\email{Contact author: simon.lahme@uni-goettingen.de}
\affiliation{Faculty of Physics, Physics Education Research, University of Göttingen, Friedrich-Hund-Platz 1, 37077 Göttingen, Germany}

\author{Dominik Dorsel}
\affiliation{Institute of Physics I and II, RWTH Aachen University, 52056 Aachen, Germany}

\author{Heidrun Heinke}
\affiliation{Institute of Physics I and II, RWTH Aachen University, 52056 Aachen, Germany}

\author{Pascal Klein}
\affiliation{Faculty of Physics, Physics Education Research, University of Göttingen, Friedrich-Hund-Platz 1, 37077 Göttingen, Germany}

\author{Andreas Müller}
\affiliation{Faculty of Sciences, Department of Physics, and Institute of Teacher Education, University of Geneva, Boulevard du Pont d'Arve 40, 1211, Genève, Switzerland}

\author{Christoph Stampfer}
\affiliation{Institute of Physics I and II, RWTH Aachen University, 52056 Aachen, Germany}

\author{Sebastian Staacks}
\affiliation{Institute of Physics I and II, RWTH Aachen University, 52056 Aachen, Germany}

\date{\today}

\begin{abstract}
This exploratory field study investigates the integration of innovative forms of recitation tasks in a first-year introductory mechanics course, focusing on smartphone-based experimental tasks alongside programming and standard recitation tasks. Smartphones, combined with external sensor modules, serve as a gateway enabling students to conduct various low-cost and authentic physics experiments with first-hand data collection outside traditional lab settings. These tasks aim to enhance students’ agency in independent physics experimentation and enrich homework assignments by dissolving boundaries between lectures, recitation sessions, and traditional labs, and thereby linking theoretical and experimental aspects of undergraduate physics education. To explore this potential, we implemented and evaluated a sample set of nine smartphone-based experimental tasks and, for comparison, three programming tasks as weekly exercises in a first-year physics course at RWTH Aachen University. We investigated students' perceptions of learning with these new tasks through twelve short surveys involving up to 188 participants, focusing on factors such as goal clarity, difficulty, or feasibility at home. In two additional surveys with 108 and 78 participants, students assessed affective responses to the smartphone-based experimental tasks relative to the programming and standard recitation tasks. Our findings indicate that the smartphone-based experimental tasks were generally well-suited to the students and tended to outperform the programming tasks in terms of perceptions of learning with the tasks and affective responses. Overall, students responded positively to the new experimental tasks, with perceptions comparable to, or only partly below, those of long-established standard recitation tasks. These results suggest that smartphone-based experimental tasks can be successfully integrated into teaching and contribute to refining traditional recitation tasks. Students’ differentiated perceptions of the three task types investigated offer valuable insights into how students perceived technology-enhanced recitation tasks in terms of feasibility, engagement, and instructional value. This provides a meaningful basis for instructors and researchers aiming to design more effective and student-centered learning environments in undergraduate physics education.
%However, students sometimes interpreted openness and real-world variability in these tasks not as opportunities, but as limitations — suggesting that prevailing norms of \glqq proper\grqq\;physics, shaped by formal instruction and assessment, can overshadow the learning potential of more exploratory formats. These insights highlight the challenges of curriculum innovation beyond the choice of tools and point to the importance of task design and framing in supporting students’ engagement with authentic, technology-supported experimentation.
%Given that most of the experimental tasks were implemented within a running course for the first time, while the standard recitation tasks have been refined over years, these results are encouraging. 
%Moreover, programming is considered an important competency also for physics students.
\end{abstract}

\maketitle
%\tableofcontents

\section{Smartphones as experimental tools}

Increasing digitalization within society has significantly transformed physics teaching and learning \cite[cf. e.g.,][]{Chen.2012,Girwidz.2019}, especially in the area of demonstration and student experiments.
Over the past 10 to 15 years, smartphones have emerged as versatile and valuable digital tools for conducting physics experiments, owing to their high prevalence and widespread availability\footnote{cf. Bring Your Own Device (BYOD) trend also in higher education \cite{Sundgren.2017}} and their various high-quality built-in sensors. Often described as "a lab in the pocket" \cite{Stampfer.2020} and "digital Swiss pocket knives for physics education" \cite{Kuhn.2022b}, smartphones support a wide range of inexpensive, qualitative and quantitative physics experiments, especially in acoustics and mechanics. They enable digitized data collection 
of different physics quantities simultaneously via multiple sensors \cite{Sukariasih.2019} with sufficient precision and accuracy similar to real labs \cite{Organtini.2022}. By replacing traditional, maybe confusing lab equipment with familiar everyday tools, smartphones may help students focus more on the experiments' content \cite{Kuhn.2022b}. Their portability makes them ideal for flexible data collection outside the classroom or lab at students' homes and in the field \cite{Vieyra.2015}. This is expected to promote authentic data collection and data sets in relevant everyday settings, following approaches of situated learning and context-based science education \cite{Vieyra.2015,Kuhn.2022b}, thereby bridging the gap between classroom and real-world problems and supporting students' motivation \cite{Kuhn.2010,Kuhn.2010b}. Smartphone-based experiments are assumed to engage students more deeply \cite{Vieyra.2015,Stampfer.2020}, particularly in building setups \cite{Organtini.2022}, and may offer autonomy in physics experimentation \cite{Kuhn.2022b} and making own decisions, which is important for open active learning \cite{Organtini.2022}. Applications like \textit{Physics Toolbox}, \textit{Science Journal}, or \textit{phyphox} make the use of built-in sensors easily accessible and include features like data export, remote access, high adaptability to the user's expertise, e.g., in data configuration and analysis \cite{Vieyra.2015,Staacks.2018}, or connectivity to external sensors \cite{Dorsel.2018}. Additionally, they provide real-time data visualization in various ways that may strengthen students' representational competence \cite{Kuhn.2022b}, graphical modeling skills \cite{Vieyra.2015}, and their ability to reflect on measurements and their alignment with models \cite{Organtini.2022}.

Due to the advantages and expectations associated with smartphones in physics education, a wide range of experiments has been developed and published in resources like the \textit{iPhysicsLabs} column in \textit{The Physics Teacher} \cite{Kuhn.2012} and specific collections \cite{Kuhn.2022,Monteiro.2022d,Organtini.2021,OBrien.2021,Sukariasih.2019,Wilhelm.2022}; a recent and very comprehensive review of smartphone-integrated physics laboratories in undergraduate physics can also be found in Ref.~\cite{Zhao.2025b}. While ample ideas for smartphone experiments exist -- especially in mechanics, making them suitable also for undergraduate physics -- surprisingly little research has investigated their effectiveness for learning physics \cite{Kuhn.2022b,Zhao.2025b}, particularly in higher education. Addressing this discrepancy, we present an approach integrating smartphone-based experimental tasks as weekly exercises in an introductory mechanics course. We assess students' perceptions of learning with these tasks as well as their affective responses -- following principles of context-based science education based on authentic tasks \cite{Kuhn.2010,Kuhn.2010b}. For comparison, we also analyze three newly implemented programming tasks and the long-established standard recitation tasks. This exploratory field study provides insights into the design, implementation, and evaluation of new technology-enhanced recitation tasks in introductory mechanics, focusing on smartphone-based experimental tasks, but also considering programming tasks as another format of contemporary physics tasks. Ultimately, this may support the development of a more engaging, student-centered study entry phase -- potentially mitigating current challenges of low retention and high dropout rates in physics programs \cite[cf.][]{Heublein.2022b,Troendle.2004}.

\section{State of research and research questions}\label{research}

\subsection{Smartphone experiments in higher physics education}

A survey among German, Finnish, and Croatian lab instructors has shown that smartphones (and tablets) are common digital tools for physics experiments in university physics labs, with 35 out of 61 instructors mentioning their use, especially during and after the COVID-19 pandemic \cite{Lahme.2023c}. During the pandemic, smartphones enabled a swift transition of lab courses to remote learning \cite{Klein.2021b,Borish.2022}. Building on these experiences, Ref.~\cite{OBrien.2021} advocates for the continued incorporation of smartphones into e-teaching physics undergraduates, particularly in labs, due to their built-in sensors and analysis tools. The author lists 75 manuscripts published between 2012 and 2020 across five journals, %\footnote{American Journal of Physics, European Journal of Physics, Physics Education, The Physics Teacher, and the proceedings of the Society of Photographic Instrumentation Engineers conferences (SPIE Optical Engineering + Applications)} 
showcasing smartphone experiments suitable for undergraduate physics education, either in on-campus labs or as at-home experiments. Similar examples of smartphone experiments suitable for higher physics education can be found also in Refs.~\cite{Kaps.2020b,Lahme.2022,Bernardini.2024,CastroPalacio.2014,Anni.2021,Ballester.2014}. Their integration into university education has already been done as preparatory or live experiments during lectures \cite{Hutz.2017,Staacks.2022}, as part of recitation exercise sheets \cite{Kaps.2021,Kaps.2022,Hutz.2017,Hutz.2019}, as undergraduate research projects \cite{Barro.2023,Lahme.2024}, and in physics labs \cite{Organtini.2022}.

\subsection{... and their evaluation}

While many ideas for smartphone experiments have been published, their use in higher physics education has rarely been evaluated. Among the 75 listed manuscripts in Ref.~\cite{OBrien.2021}, only 8 manuscripts (11\%) report any form of evaluation involving learners. Reported evaluation findings include successful implementation of an experimental task \cite{Salinas.2018}, an unproblematic workflow \cite{Pörn.2016}, or the perception of smartphone application as a convenient alternative to standard lab equipment offering immediate visualization of experimental data \cite{Hawley.2018}. They noted increased student interest, motivation, and enthusiasm \cite{Hawley.2018,Rayner.2017,Testoni.2016,Shakur.2016}, a better understanding of physics content \cite{Rayner.2017,Arribas.2015,Testoni.2016}, or more reflective quantification of constants \cite{Azhikannickal.2019}. There were also impressions that students worked more exploratory \cite{Pörn.2016}, were proud of their self-generated data \cite{Shakur.2016}, and became more comfortable with experimental designs \cite{Arribas.2015}. While the authors were typically not expected to present evaluation data in their manuscripts focusing on their experimental task idea, the rare and then rather superficial evaluation, partly just based on anecdotal evidence, highlights the need for more systematic evaluation of smartphone experiments in university physics education. As far as we know, more in-depth evaluation research on the use of smartphone sensors for physics experiments in this context has only been carried out by Refs.~\cite{Organtini.2022,Bernardini.2024,Kaps.2021,Kaps.2022}, briefly summarized below. 

Reference~\cite{Organtini.2022} integrated five experiments using smartphones or Arduinos, in addition to one to two experiments using dedicated devices, into a physics lab for Italian physics undergraduates during the COVID-19 pandemic. The evaluation used end-of-course questionnaires and the E-CLASS instrument \cite{Zwickl.2014} assessing students' beliefs about their own and experts' abilities in experimental physics within a lab course context. The authors report that their course was not evaluated very positively. A significant gain in the E-CLASS score occurred only in the first of the two years studied; there were no significant differences between subgroups of students who used smartphones and Arduinos more frequently compared to those who used more standard lab equipment. The authors discuss these outcomes by pointing to some poor decisions in course design and referencing Ref.~\cite{Deslauriers.2019}, which showed "that self-reported perception of learning of students is, on average, lower than their actual learning" (p.~7). They note that their implementation of Arduinos and smartphones at least did not harm students, while it benefited high-performing students. Although the study contributes valuable information, its interpretability would benefit from the inclusion of key descriptive statistics, such as sample sizes, as well as a more detailed breakdown of groups and settings under comparison.

Reference~\cite{Bernardini.2024} developed, piloted, and partly refined two smartphone experiments for a first-year physics course for engineering, architecture technology, or computer science students at an American and Italian university. In the first task, students investigated the period duration of a pendulum using the proximity sensor placed below the pendulum as a stopwatch. In the second task, a bouncing ball was studied by deriving data from the preset \textit{phyphox}-configuration \textit{acoustic stopwatch}. The tasks were supplemented with supportive materials like video tutorials, guiding questions, and Excel workbooks. The evaluation included short student questionnaires, instructors' observations, and rubrics on students' learning outcomes, although not all outcomes were reported in the manuscript. Regarding the pendulum task, students liked the new and easy mode of data collection and the application of theory to real data. They disliked the lengthy data collection and analysis and reported difficulties with the spreadsheet and sensor functionality, requesting more detailed explanations. The authors concluded that "smartphone-based laboratory experiences can be successfully used in introductory physics experiences, even in university settings" but that "it is critical to reflect on the support to be provided to the students and on the instructional actions to be put in place in order to achieve the desired learning outcomes" (p.~9).

Reference~\cite{Kaps.2021} reports an initial evaluation of smartphone-based experimental tasks for undergraduate German physics teacher training students. According to the students, these tasks were engaging and fostered interest in learning more about the physics content, though they were also time-consuming. Compared to a previous cohort that used only paper-pencil versions of the tasks, the performance of the cohort using smartphones was lower, which the authors attribute to increased task complexity. However, pilots showed a medium-sized effect on students' understanding and application of physics concepts within smartphone-based tasks. In a follow-up study \cite{Kaps.2022}, a quasi-experimental field study was conducted with two groups of first-semester students: an investigation group of 40~first-semester teacher training students who conducted smartphone experiments to evaluate their physics models, and a control group of 45~physics and meteorology students who worked on analog but algebra- and calculus-based tasks. In a pre-post-design, neither initial group differences nor changes over time were found in both groups regarding students' motivation and curiosity. The authors suggested that this may be explained by the diminished novelty effect of smartphones until the post-test and the pressure associated with the tasks being part of exam prerequisites. However, it should be noted that the circumstances of the intervention and control groups are rather difficult to compare, which likely influenced the study's results.

Beyond these studies from higher education, investigations with pupils show little differences between groups working with smartphones and those with comparable tasks without smartphones. Ref.~\cite{Mazzella.2016} found no difference in conceptual understanding of acceleration and dealing with vector representations; the smartphone group was only better in designing an experiment to measure acceleration and describing acceleration in a free fall motion. Similarly, Ref.~\cite{Hochberg.2018} found no differences in learning achievement but positive effects on pupils' interest and topic-specific curiosity in the smartphone group.

All in all, the state of research shows that, despite the broad development of smartphone experiments, there has still been rather low interest in the implementation and research-based evaluation of smartphone experiments in university physics education. Existing studies often use not established or invalidated evaluation instruments, lack sophisticated analysis, or show limited comparability of treatment and control groups.

\subsection{Research questions}

To address the current lack of empirical, evaluative research on the implementation and effectiveness of smartphone-based experimental tasks in university physics education, we conducted an exploratory field study in a first-year mechanics course. We designed, implemented, and systematically evaluated a sample set of smartphone-based experimental tasks that were used as weekly exercises. Following the approach of context-based science education \cite{Kuhn.2010,Kuhn.2010b}, the study focuses on students’ perceptions of learning with the tasks and their affective responses, while also exemplifying how such evaluations of new technology-enhanced teaching approaches can be meaningfully embedded into a running course.

In order to interpret the results more robustly, we compared these new experimental tasks to two other task types that were also implemented in the same course. First, we included multiple standard recitation tasks, i.e., established textbook tasks that are frequently used as weekly exercises throughout the course. These long-established tasks serve as an essential reference point for evaluating the added value and distinctiveness of newer formats in introductory physics courses. Second, we included three Python-based programming tasks that were newly introduced in the same course. These tasks were introduced as part of a new faculty-led teaching approach in alignment with the growing emphasis on integrating computational thinking and programming into standard physics curricula \cite[cf.][]{Atherton.2023,Chabay.2008}. They also served a preparatory role for two of the experimental tasks, which required Python for data analysis. Thus, a parallel evaluation and comparison of experimental and programming tasks allows for a deeper assessment of the smartphone-based experimental tasks, while also providing first insights into programming tasks as another important and innovative format of technology-enhanced tasks in undergraduate physics education.

Overall, our study addresses three research questions aimed at evaluating the smartphone-based experimental tasks in comparison with the established standard recitation tasks and the newly introduced programming tasks.\\
RQ1: \textit{How do students perceive learning with the smartphone-based experimental tasks in comparison to the programming tasks?}\\
RQ2: \textit{What are students' affective responses towards the experimental tasks in comparison to the programming and standard recitation tasks?}\\
RQ3: \textit{How do potential predictors (e.g., gender, study program) influence perception of learning with the tasks and affective responses?}

The focus on students' perceptions of learning with the tasks and affective responses, which include relevant target variables both of which are of interest regarding the use of smartphone experiments (cf. state of research), is appropriate for all three task types despite partly different learning goals. The comparison provides deep insights into students' perceptions of smartphone-based experimental tasks in undergraduate physics compared to two other important task types. 

\section{Materials and methods}

\subsection{Implemented tasks}\label{tasks}

\begin{figure*}[htb]
    \centering
    \includegraphics[width=.8\textwidth]{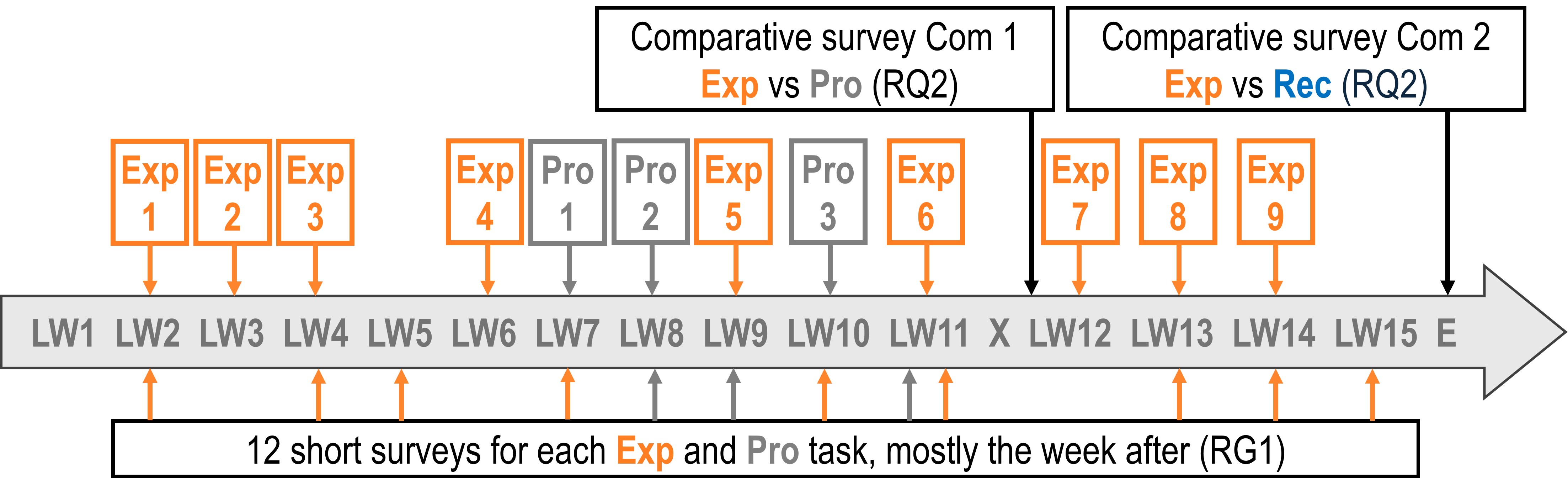}
    \vspace{-0.35cm}\caption{Overview of the implementation and evaluation of nine smartphone-based experimental and three programming tasks (cf. Table~\ref{tab:overviewtasks} in the Appendix) across 15 lecture weeks (LW). Each task was evaluated with a short survey mostly the week after, marked by arrows below the timeline. In LW12 and the first exam week (E) after lecture time, two additional comparative evaluations Com~1 and Com~2 were conducted to compare the experimental tasks with the programming tasks (Exp vs. Pro) and the standard recitation tasks (Exp vs. Rec) respectively.}% Exp~1 was the only task that students conducted on-campus within their recitation group; all other tasks were homework exercises.}
    \label{fig:design}
\end{figure*}

As shown in Fig.~\ref{fig:design}, nine smartphone-based experimental tasks (Exp) and three programming tasks (Pro) were integrated into weekly exercise sheets of the introductory physics course \textit{Experimental Physics I} at RWTH Aachen University during the winter semester~23/24. This introductory mechanics course primarily serves over 300 first-year physics majors and future high school teachers. The course consists of two weekly lectures and weekly recitation groups. Students typically have a physics lab course not before the end of their second semester. Of the 13 weekly exercise sheets, with about 5 tasks each, 12 included one Exp- or Pro-task. Of these tasks (cf. Table~\ref{tab:overviewtasks} in the Appendix), seven tasks (four Exp- and all three Pro-tasks) were implemented into teaching for the first time. Two Exp-task have been used already in a prior cohort but have been significantly modified, and the other three Exp-tasks have already been used in teaching in the same way before. The remaining tasks were standard recitation tasks (Rec; cf. Table~\ref{Tab:TaskTypes} for a comparison of the three task types). Students worked self-directed on all tasks in groups of three at home, except for the first Exp-task, which was done guided during the first recitation session. Upon completion, students submitted their work to a tutor for correction and grading. Each exercise sheet was assigned 20 points; in total were 260~points possible, including 50~points (19\%) from Exp-tasks and 15~points (6\%) from Pro-tasks. To qualify for the final written exam, students needed to earn at least half of the possible points.

\begin{table*}[htb]
\caption{Characteristics of the three task types compared in this study.}
\begin{ruledtabular}
\footnotesize
\begin{tabular}{p{.17\textwidth}p{.32\textwidth} p{.235\textwidth} p{.235\textwidth}}
&Experimental tasks (Exp)&Programming tasks (Pro)&Standard recitation tasks (Rec)\\\hline
Main focus&\hangindent=.3cm Hands-on experimentation, partly involving the derivation and application of physics formulas&\hangindent=.3cm Completing pre-written code&\hangindent=.3cm Traditional problem-solving exercises\\
Physics topics&\hangindent=.3cm Kinematics, dynamics, rotating and rolling motions, oscillations&\hangindent=.3cm Path integral, relativistic mechanics, Kepler's law&\hangindent=.3cm Across all lecture topics throughout the semester\\
Technologies&\hangindent=.3cm Smartphones (\textit{phyphox}), partly external sensor boxes, partly Excel or Python for data analysis&\hangindent=.3cm Python scripts&None (pen and paper)\\
Level of guidance&\hangindent=.3cm Detailed, partly illustrated step-by-step instructions, pre-set configuration of \textit{phyphox}&\hangindent=.3cm Pre-written and annotated Jupyter notebooks&\hangindent=.3cm Standard, rather closed task instructions\\
Required skills&\hangindent=.3cm Designing and conducting experiments; collecting, analyzing, interpreting, and visualizing data&\hangindent=.3cm Understanding and modifying given code by adding derived formulas&\hangindent=.3cm Deriving formulas, doing calculations, drawing sketches/diagrams\\
Student submission&\hangindent=.3cm Photos of experimental setups and results of data analysis&\hangindent=.3cm Completed Jupyter notebooks&\hangindent=.3cm Written solutions, including formulas, figures, sketches, or diagrams\\
\hangindent=.3cm Quantity and percentage of exam prerequisite points&\hangindent=.3cm 9 tasks across the semester (19\%)&\hangindent=.3cm 3 tasks in the middle of the semester (6\%)&\hangindent=.3cm Around 50, 4 to 5 tasks per week (75\%)\\
Exam relevance&Low&Low&Mostly high to very high\\
Level of novelty&Mostly used for the first time (in this way)&Used for the first time&Established, commonly used\\
\end{tabular}
\end{ruledtabular}
\label{Tab:TaskTypes}
\end{table*}

\subsubsection{Experimental tasks (Exp)}

The experimental tasks (Exp) use smartphones equipped with the \textit{phyphox} application and various built-in sensors, as well as external sensor boxes lent to the students. For most tasks, students received a preset configuration of \textit{phyphox}, enabling task-specific data collection and visualization, thereby simplifying the application's use. Additional materials were either common household items like paper or aluminum foil, which students supplied themselves, or specific materials such as strings or golf balls, which were lent to them.

Each task included detailed, partially illustrated step-by-step instructions for using the application, setting up the experiment, and typically also for data collection, covering the procedure, measurement plan, and key steps of data analysis. Common data collection errors and safety concerns were explicitly addressed to help students avoid gathering incorrect data, injuring themselves, or damaging equipment. 

The complexity of data analysis varied across tasks, from only reporting derived measures automatically calculated by the application to larger data processing and visualization using external tools like Excel or Python. The final two tasks, Exp~8 and Exp~9, involved more extensive data analysis, processing, and visualization in Python, with self-written code in Exp~8 and a given Jupyter notebook in Exp~9. To verify their work, students had to submit a photo of their experimental setup along with the results of their data analysis.

Before implementation, each smartphone experiment was thoroughly tested by at least one person from the \textit{phyphox} group to ensure its experimental and technical feasibility. The experimental ideas, for which there was already a wealth of experience within the group of authors, were then translated into task instructions. The instructions were iterated within a subgroup of authors with diverse backgrounds in physics, physics education research, and undergraduate physics instruction. This ensured that the tasks were of sufficient quality for their first use in our study.

\begin{figure}
    \centering
    \includegraphics[width=0.73\columnwidth]{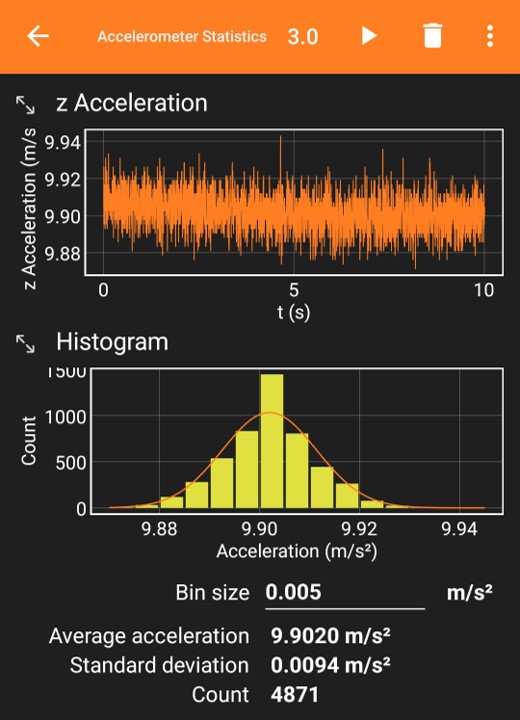}
    \vspace{-0.3cm}\caption{Snippet of task Exp~2: Screenshot from \textit{phyphox} when determining mean and standard deviation of the noise data of the acceleration sensor for different measurement periods and tilting angles (example data for a duration of 10~s and a tilting angle of 0°).}
    \label{fig:Exp2}
\end{figure}

Three Exp-tasks are outlined in more detail in the following. In Exp~2 (Sensor statistics of the smartphone), students analyze the measurement uncertainty of their smartphones' acceleration sensor in the $z$-direction. Using the preset \textit{phyphox}-configuration \textit{Accelerometer Statistics} (cf. Fig.~\ref{fig:Exp2}), sensor data is collected over a set time interval. Students determine and compare the mean, standard deviation, and uncertainty of the mean for the noise data across three different measurement periods and two different tilt angles of the smartphone. Additionally, students evaluate the provided statistics from two other smartphones. Exp~2 is the only task that requires no equipment beyond the smartphone and directly addresses measurement uncertainties; none of the other tasks force a quantitative assessment of uncertainties.

\begin{figure}
    \centering
    \includegraphics[width=\columnwidth]{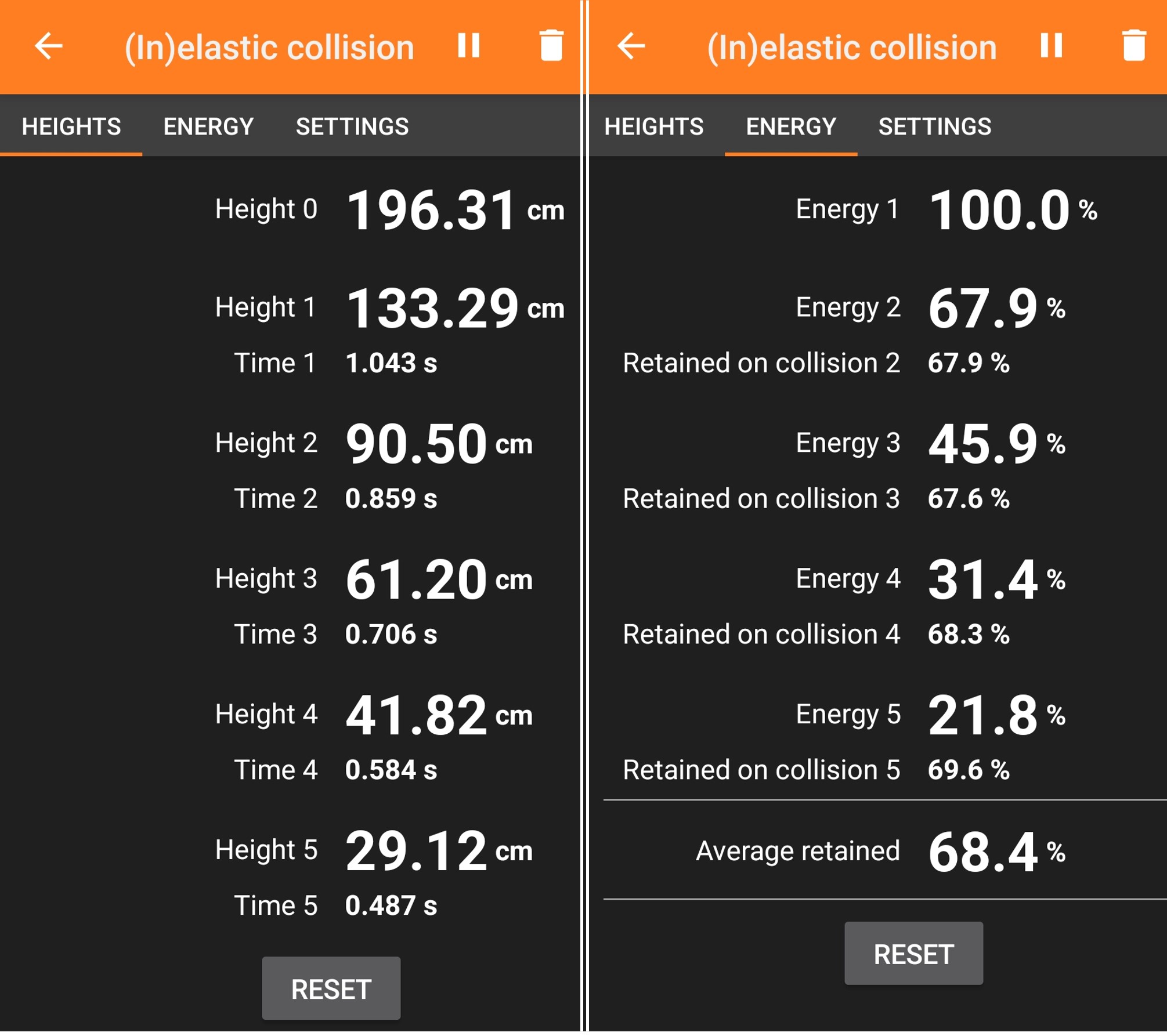}
    \vspace{-0.7cm}\caption{Snippet of task Exp~5: Screenshots from \textit{phyphox} when measuring time intervals of a bouncing ball based on acoustic signals. By this, students analyze, among others, the percentage of conserved energy per bounce for different damping grounds.}
    \label{fig:Exp5}
\end{figure}

In Exp~5 (Bouncing ball, cf. Appendix for task instructions), students study the behavior of bouncing golf and table tennis balls on different layers of damping paper to determine the initial drop height and analyze the percentage of conserved energy after each bounce. Using the preset \textit{phyphox}-configuration \textit{(In)elastic collision} (cf. Fig.~\ref{fig:Exp5}), students acoustically measure the intervals between the ball's bounces, with the application automatically calculating the maximum height reached between consecutive bounces. They conduct the experiments, derive equations to model the process, compare experimental findings with their predictions, and plot the fraction of energy conserved during the second bounce for both balls as a function of the number of damping paper layers used. This task exemplifies the integration of experimental and theoretical subtasks, as in other tasks like Exp~2 and Exp~4, and demonstrates how students can conduct smartphone experiments just with lent everyday objects.

\begin{figure}
    \centering
    \includegraphics[trim={0 1.6cm 0 0}, clip, width=\columnwidth]{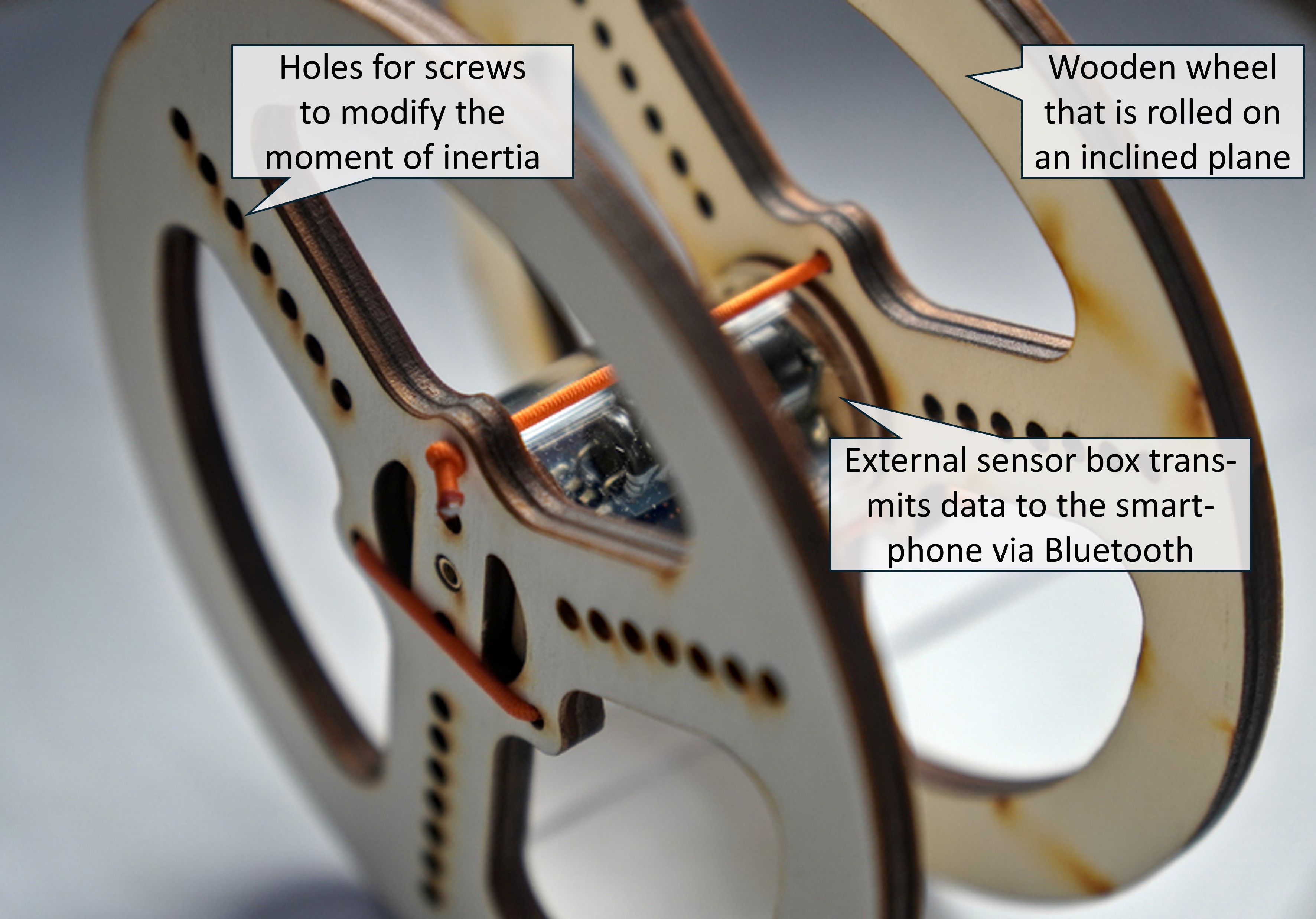}
    \vspace{-0.7cm}\caption{Snippet of task Exp~6: Students investigated rolling motions of a wooden wheel on an inclined plane with an external sensor box (gyroscope). By attaching screws and washers the moment of inertia is varied, which should be determined based on angular velocity data.}
    \label{fig:Exp6}
\end{figure}

In Exp~6 (Moment of inertia on an inclined plane), students modify and determine the moment of inertia of a wooden wheel on an inclined plane measuring angular velocity over time. They were lent various equipment, including a wooden wheel, screws, washers, and an external sensor box equipped with multiple sensors, including a gyroscope that connects to the smartphone via Bluetooth (cf. Fig.~\ref{fig:Exp6}). To systematically vary the moment of inertia, students attach screws and washers at five different distances from the rotational axis. By rolling the wheel down an inclined plane, they experimentally determine the moment of inertia using angular velocity data. They finally plot the moment of inertia in dependency on the distance between the attached weights and the rotational axis.

\subsubsection{Programming tasks (Pro)}\label{Sec:Pro}

The three programming tasks (Pro) are structured by Jupyter notebooks that offer extensive pre-written and annotated code. This high level of guidance allows students to focus on physics concepts and applying physics formulas within the provided code, rather than getting caught up in technical programming details. While programming is not taught in the course itself, most students were required to take a concurrent programming course, to which the Pro-tasks were tailored. For students who did not take this course due to other study programs, these were optional. As for the Exp-tasks, the instructions of the Pro-tasks were iterated within a subgroup of authors.

\begin{figure*}
    \centering
    \includegraphics[width=0.83\textwidth]{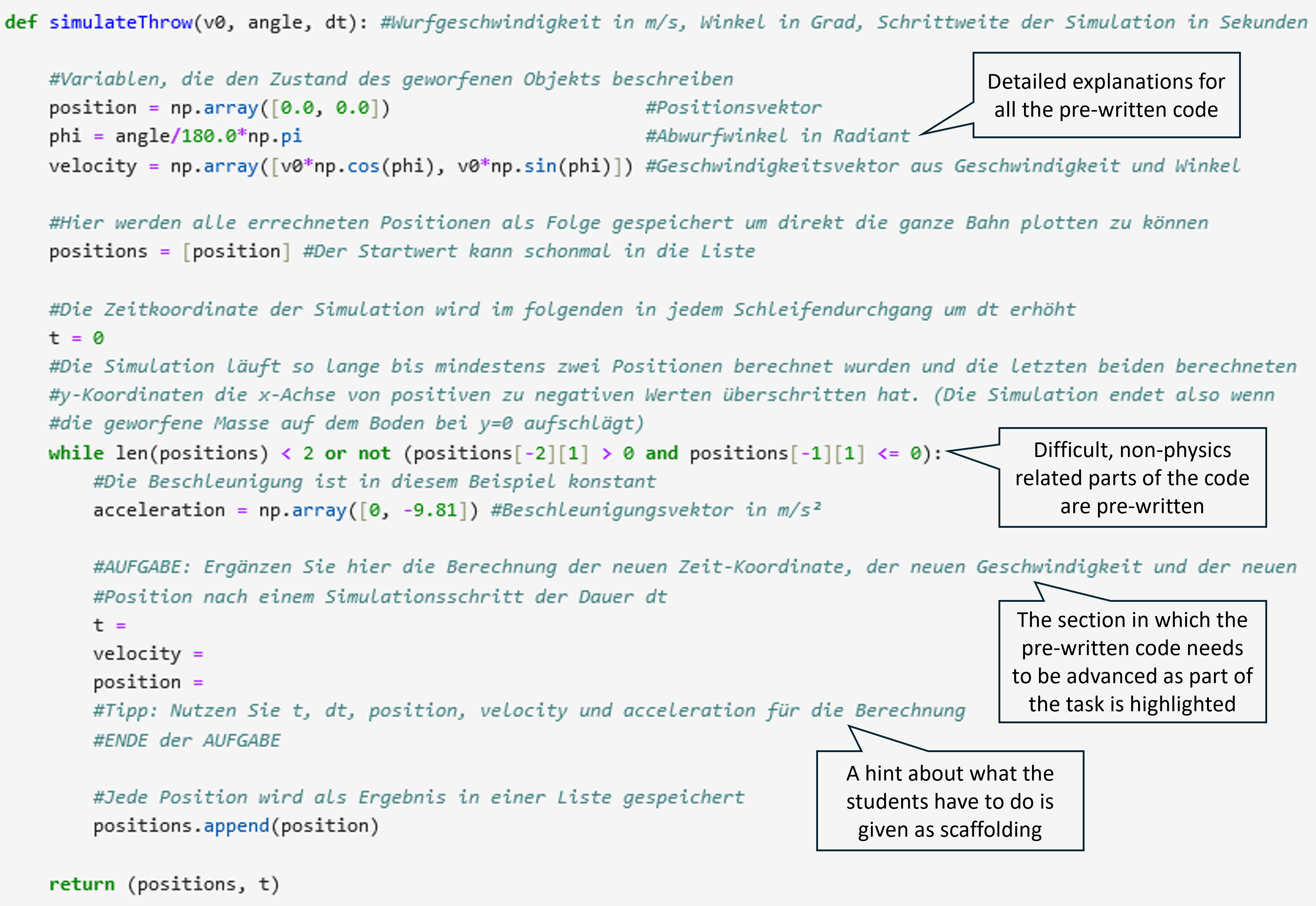}
    \vspace{-0.3cm}\caption{Snippet of the Python script given to the students as part of Pro~3 (Simulated third Kepler's law). Larger code parts are pre-written. Students had to define time, velocity, and position of an object after incremental time when thrown to  earth at a given angle with initial velocity.}
    \label{fig:Pro3}
\end{figure*}

The Pro-tasks facilitate topics that go beyond the possibilities of smartphone experiments. For instance, in task Pro~3, students simulate planetary orbits. In the first subtask (cf. Fig.~\ref{fig:Pro3}), they define the function \textit{simulateThrow} that returns a list of positions and times of an object thrown at the Earth with an initial velocity $v_0$ and an initial angle $\varphi$. As most of the code is already provided, students only need to code the formulas for the time, velocity, and position of the object after a time increment $dt$. In the second subtask, students create a similar function returning the list of positions and times for a planet, based on its initial velocity and distance to the Sun. Students reuse their code from the first subtask and write the formula for the acceleration due to the Sun at the center of the coordinate system. In the final subtask, students must define another function that calculates the semi-major axis of a planet using the position and time list. The code then outputs the orbital period $T$, the semi-major axis $a$, and the ratio $\frac{T^2}{a^3}$, enabling students to compare the simulation results with real parameters.

\subsubsection{Standard recitation tasks (Rec)}\label{Sec:Rec}

The standard recitation tasks (Rec) make up the majority of tasks that students complete as part of their weekly exercise sheets. They are common textbook problems used for exercise sheets in physics courses at many universities. According to the taxonomy of physics problems by Ref.~\cite{Teodorescu.2013}, they are mostly either retrieval tasks involving the execution of mental procedures or comprehension tasks symbolizing physics mental procedures. The retrieval tasks involve "performing a procedure or task needed to solve the problem without significant error [...]. The student can calculate and compute different physical quantities, derive physics results, draw free-body diagrams, graph, plot physical quantities, and solve physics-related equations. [...] [T]he students need to know how to perform the mental procedures particular to physics problem solving (e.g., solve mathematical equations, calculate different physical quantities using their definition, and so on)" (pp.~7f). The comprehension tasks involve the construction of "an accurate symbolic image of the information or mental procedure needed to solve the physics problem. [...] The student can represent pictorially, verbally, mathematically, or graphically physical quantities, basic equations, formulas, physics vocabulary terms, concepts, symbols, and phenomena" (p.~8). The Rec-tasks are established tasks that have been similarly used for several years and are representative of those found on the final written exam.

\subsection{Data collection and samples}

Perceptions of learning with the tasks and affective responses were evaluated in 14~online surveys (cf. Fig.~\ref{fig:design}). Twelve short surveys evaluated every Exp- and Pro-task in the weekly recitation groups, mostly the week after the task was completed. After all Pro-tasks but before the Exp-tasks requiring Python for data analysis, a longer survey was used to compare both task types. A similar survey comparing Exp- and Rec-tasks was conducted the week after the lecture time.

Pseudonym codes were used to link participants' responses across all surveys. Codes with potential misspellings were matched if at least three out of four code elements were identical, totaling in 313~codes. Study participation, including partial responses, varied from 188~students for Exp~1 to 41~students for Exp~9; the average participation was $M=94$ ($\mathrm{SD}=48$). The comparative surveys Com~1 and Com~2 were done by 108 and 78~students, respectively, whereby 3 participants from Com~1 and 4 participants from Com~2 were excluded for not conscientious answering behavior.

In the first survey, all 188 students provided demographic data. 50~participants identified as female, 126 as male, and 4 identified as non-binary; 8 did not specify a gender. 129~students were enrolled in the standard physics bachelor, 33 in a special four-year physics bachelor, 20 in a physics teacher training program, and 6 in another program. 156 students were in their first semester; the average semester was $M=1.6$, $\mathrm{SD}=1.9$. 182 students also reported their final school grade, which was $M=1.8$, $\mathrm{SD}=0.6$ on a scale where $1.0$ is the best grade and $4.0$ is the minimum grade for graduation. \textit{Prior experience with physics experiments} in general \textit{and smartphone experiments} in particular varied, ranging from students with no prior experience at all to students who have already conducted multiple of these experiments in school or beyond (cf. Table~\ref{tab:priorexperience} in the Appendix).

\subsection{Instruments}

Table~\ref{Tab:Variables} displays the variables measured in the surveys for each task and the two comparative surveys according to the research questions. All items are presented numbered in German and with English translations as supplementary material.

\begin{table*}[htb]
\caption{Overview of the investigated variables indicated by an "X" that were used in the weekly surveys for the experimental (Exp) and programming (Pro) tasks and the comparative surveys (Com) for the experimental, programming, and standard recitation (Rec) tasks.}
\begin{ruledtabular}
\footnotesize
\begin{tabular}{p{.275\textwidth}p{.05\textwidth}p{.04\textwidth}p{.36\textwidth}p{.07\textwidth}p{.075\textwidth}p{.08\textwidth}}
Variable&based on Ref.&$N$ items&Scale&Exp~1-9\newline\& Pro~1-3&Com~1\newline Exp vs. Pro&Com~2\newline Exp vs. Rec\\\hline
\multicolumn{3}{l}{\textbf{Students' perceptions of learning with the tasks (RQ1)}}&&&&\\
Overall task rating&\cite{Lahme.2023b}&1&metrical, 1 (worst) - 10 (best)&X&&\\
Task quality&\cite{Lahme.2023b}&7&metrical, 1 (strongly disagree) - 5 (strongly agree)&X&&\\
Task difficulty&\cite{Lahme.2023b}&1&ordinal (too easy - adequately challenging - too difficult)&X&&\\
Sufficiency of instructions&\cite{Lahme.2023b}&1&ordinal (too detailed - sufficient - insufficient)&X&&\\
Time spent on task&\cite{Lahme.2023b}&1&open numerical input&X&&\\
Contribution to the task&\cite{Rauschenbach.2018}&1&ordinal&X&&\\
Use of technologies&\cite{Lahme.2023b}&6&metrical, 1 (strongly disagree) - 5 (strongly agree)&&X&\\
Open task feedback&\cite{Lahme.2023b}&3&open-text field&X&&\\
\textbf{Affective responses (RQ2)}&&&&&&\\
Curiosity&\cite{Klein.2016}&8&metrical, 1 (doesn't apply at all) - 6 (is completely true)&&X&X\\
Interest&\cite{Klein.2016}&16&metrical, 1 (doesn't apply at all) - 6 (is completely true)&&X&X\\
Authenticity&\cite{Klein.2016}&18&metrical, 1 (doesn't apply at all) - 6 (is completely true)&&X&X\\
Experience during the tasks&\cite{Lahme.2023b}&12&metrical, 1 (strongly disagree) - 5 (strongly agree)&&X&X\\
Perceived educational effectiveness&\cite{Lahme.2023b}&6&metrical, 1 (strongly disagree) - 5 (strongly agree)&&&X\\
Autonomy&\cite{Klein.2016}&4&metrical, 1 (doesn't apply at all) - 6 (is completely true)&&&X\\
Linking to the lecture&\cite{Rehfeldt.2017}&3&metrical, 1 (strongly disagree) - 6 (strongly agree)&&&X\\
\textbf{Potential predictors (RQ3)}&&&&&&\\
Gender&&1&nominal&Exp1&&\\
Semester&&1&open numerical input&Exp1&&\\
Study program&&1&nominal&Exp1&&X\\
High school graduation grade&&1&open numerical input&Exp1&&\\
\raggedright\hangindent=.3cm Prior experience 
physics experiments&&1&open-text field, ordinally categorized&Exp1&&\\
\raggedright\hangindent=.3cm Prior experience 
smartphone experiments&&1&open-text field, ordinally categorized&Exp1&&\\
Affordances&\cite{Fliege.2001}&5&metrical, 1 (almost never) - 6 (most of the time)&X&X&\\
Sense of belonging to university&\cite{Baumert.2008}&6&metrical, 1 (not true at all) - 4 (quite true)&Exp2,9&&\\
Sense of belonging to physics community&\cite{Feser.2023c}&1&metrical, 5 specific answer options&Exp2,9&&\\
Academic self-concept&\cite{Dweck.1999,Dresel.2013}&3&metrical, 6-point scale with specific statements&Exp1,8&&\\
\end{tabular}
\end{ruledtabular}
\label{Tab:Variables}
\end{table*}

The weekly surveys addressed the students' perceptions of learning with each Exp- and Pro-task (cf. RQ1). Students gave the tasks an \textit{overall task rating} and provided feedback in open-text field questions on what they liked and disliked about the task and what they would like to change. Other items addressed the perceived \textit{task quality}, i.e., the clarity of learning goals and task instructions, the design of task instructions, the sufficiency of instructions on the technologies used, the relationship to the field of study, and whether requirements are given and students feel confident to complete the task at home. Two items addressed \textit{task difficulty} and \textit{sufficiency of instructions} and supporting materials. Students also estimated the \textit{time spent on task} in an open-text field and reported how much they contributed to the discussion and task completion within their group. All items in the weekly surveys were the same for both types of tasks, with only the specification "experimental task" or "programming task". Facing question RQ3, these surveys also measured ten possible predictors (cf. Table~\ref{Tab:Variables}).

In the two comparative surveys, students compared the Exp-tasks with the Pro-tasks (Com~1) and Rec-tasks (Com~2) regarding their affective responses (cf. RQ2). Some variables were asked in both surveys to compare all three types of tasks, while others were only included in one survey, depending on their relevance and constraints of survey length. In both surveys, students rated items about the \textit{curiosity} and \textit{interest} generated by the tasks, perceived task \textit{authenticity}, and the \textit{experience during the tasks}. In addition, the first comparative survey asked about the use of digital technologies in the tasks, while the second one asked for the perceived educational effectiveness, perceived \textit{autonomy}, and \textit{linking to the lecture}. Every item was presented twice in a row for both task types to facilitate direct comparison and readability. The first comparative survey was written in the present tense and the second one in the past tense to account for the students' progress in the semester between the measurement times.

\subsection{Instrument analysis}\label{instrumentanalysis}

\begin{table*}[htb]
\caption{Results of the exploratory factor analysis for variables used to evaluate the experimental tasks. For factors with eigenvalues $>1$ and an explained variance $\geq 10\%$, the number $N$ of items and cases, eigenvalues, and explained variance $\mathrm{Var}$ are given. Minimal and maximal factor loads show the range of correlations between items and factors (in case of more than one factor after varimax rotation). Further, descriptive data (minimum $\mathrm{Min}$, maximum $\mathrm{Max}$, mean $M$, and standard deviation $\mathrm{SD}$), Cronbach's $\alpha$, the maximum possible Cronbach's $\alpha_{max}$ if an item is deleted, the minimal item-total correlation $r_{i(t-i)}$, and the minimal and maximal item difficulties $p_{min}$ and $p_{max}$ are given. * indicates a 5-point Likert scale, otherwise, it is a 6-point Likert scale. Values in \textit{italics} exceed or undermine recommended thresholds.}
\begin{ruledtabular}
\footnotesize
\begin{tabular}{p{.29\textwidth}p{.035\textwidth}p{.035\textwidth}p{.04\textwidth}p{.035\textwidth}p{.055\textwidth}p{.055\textwidth}p{.035\textwidth}p{.035\textwidth}p{.035\textwidth}p{.035\textwidth}p{.035\textwidth}p{.035\textwidth}p{.04\textwidth}p{.035\textwidth}p{.035\textwidth}}
Scales (\& included items, cf. supplementary material)&$N$ items&$N$ cases&Eigen-value&Expl. $\mathrm{Var}$&Min.\,fac-tor load&Max.\,fac-tor load&$\mathrm{Min}$&$\mathrm{Max}$&$M$&$\mathrm{SD}$&$\alpha$&$\alpha_{max}$&Min. $r_{i(t-i)}$&$p_{min}$&$p_{max}$\\\hline
\multicolumn{4}{l}{\textbf{Students' perceptions of learning with the tasks (RQ1)}}&&&&&&&&&&&&\\
Task quality* (Qua1-7)&7&162&&&&&1.00&5.00&4.17&0.61&0.86&0.85&0.56&0.77&0.90\\
$\mapsto$ Goal clarity* (Qua1-3,5)&4&162&3.79&0.54&0.70&0.84&1.00&5.00&4.02&0.67&0.81&0.77&0.60&0.78&0.82\\
$\mapsto$ Feasibility at home* (Qua4,6,7)&3&162&1.18&0.17&0.79&0.90&1.00&5.00&4.36&0.74&0.87&0.85&0.70&0.84&0.90\\
Use of technologies* (UoT1-6)&6&87&3.14&0.52&0.58&0.78&2.00&4.83&3.71&0.63&0.81&0.82&0.43&0.59&0.76\\
\textbf{Affective responses (RQ2)}&&&&&&&&&&&&&&&\\
Curiosity (Cur1-8)&8&105&4.47&0.56&0.55&0.83&1.63&5.88&3.61&0.91&0.88&0.89&0.45&0.45&0.65\\
Interest (Int1-16)&16&101&7.49&0.47&0.38&0.85&1.61&5.44&3.61&0.90&0.92&0.91&0.35&0.37&0.80\\
Authenticity (Auth1-18)&18&92&&&&&2.50&6.00&4.16&0.83&0.93&0.93&0.47&0.51&\textit{0.92}\\
$\mapsto$ Disciplinary authenticity (Auth10-18)&9&92&8.23&0.46&0.53&0.80&2.00&6.00&4.16&0.83&0.90&0.90&0.52&0.57&0.89\\
$\mapsto$ Reference to reality (Auth1-9)&9&92&1.85&0.10&0.50&0.86&2.00&6.00&4.16&0.80&0.88&0.88&0.54&0.51&\textit{0.92}\\
Experience during the tasks* (EdT1-12)&12&98&&&&&2.33&4.50&3.40&0.50&0.72&0.75&\textit{0.03}&\textit{0.28}&0.74\\
$\mapsto$ Experience of competence* (EdT1,7-9)&4&98&3.43&0.34&0.75&0.84&1.00&5.00&3.45&0.86&0.82&0.81&0.54&0.45&0.74\\
$\mapsto$ Curiosity/Interest (short)* (EdT2,5,6,10)&4&98&2.00&0.20&0.56&0.88&1.75&5.00&3.46&0.73&0.75&0.78&0.39&0.33&0.71\\
$\mapsto$ Autonomy/Creativity (short)* (EdT11,12)&2&98&1.19&0.12&0.58&0.92&1.50&5.00&3.39&0.80&-%\textit{0.45} Cronbachs'alpha not meaningful, one would have to calculate Spearman-Brown-coefficient, if desired
&-&0.30&0.52&0.62\\
%\textbf{RG2b}&&&&&&&&&&&&&&&\\
Perceived educational effect.* (Eff1-6)&6&74&&&&&1.86&4.57&3.62&0.62&0.80&0.80&0.41&0.41&\textit{0.92}\\
$\mapsto$ Perceived affective effect.* (Eff4-6)&3&74&3.05&0.51&0.81&0.82&1.33&5.00&3.39&0.87&0.79&0.75&0.60&0.41&0.68\\
$\mapsto$ Perceived cognitive effect.* (Eff1-3)&3&74&1.20&0.20&0.70&0.88&2.00&5.00&3.95&0.64&0.75&0.69&0.59&0.53&\textit{0.92}\\
Autonomy (Aut1-4)&4&74&&&&&1.50&5.50&3.91&0.85&0.66&0.84&\textit{0.04}&0.38&0.74\\
$\mapsto$ Autonomy (Aut1,3,4)&3&74&2.28&0.76&0.85&0.88&1.00&6.00&4.15&1.04&0.84&0.81&0.67&0.72&0.74\\
Linking to the lecture (Lin1-3)&3&66&1.89&0.63&0.79&0.80&1.00&6.00&4.43&0.88&0.70&0.62&0.52&0.65&\textit{0.94}\\
\end{tabular}
\end{ruledtabular}
\label{tab:factoranalysis}
\end{table*}

We applied an exploratory factor analysis with varimax rotation on all variables from the comparative surveys and the \textit{task quality} items from the weekly surveys. The responses from Exp-tasks served as datasets, as they are the main focus of interest. Whenever possible, data from Com~1 %(e.g., for \textit{use of technologies})
were used, given its higher participation rate; otherwise, data from Com~2 %(e.g., for \textit{perceived educational effectiveness}) 
were used. For \textit{task quality}, processed data from survey Exp~2 were used, as it had the second-highest participation rate after Exp~1 and because Exp~1 was deemed unrepresentative being the only task conducted on-campus, i.e., not at home.% The few inverted items were reinverted.

All variables met the prerequisites for factor analysis. We considered only factors with a minimum eigenvalue of 1.00 and an explained variance of at least 10\%. The number of factors was determined based on these criteria and the scree plot. To identify which items belonged to each factor, we used the rotated component matrix. For each factor, Cronbach's $\alpha$ was checked to ensure it is satisfying (between 0.65 and 0.95) and whether excluding an item would noticeably increase Cronbach's $\alpha$. Furthermore, for each item within the factors, we checked whether the minimum item-total correlation exceeded the recommended threshold of 0.30 and whether the percentage of students agreeing with a statement (item difficulty $p$) fell within the recommended range of 0.30 to 0.90.

The results (cf. Table~\ref{tab:factoranalysis}) indicate that \textit{use of technologies}, \textit{curiosity}, \textit{interest}, and \textit{linking to the lecture} are satisfactory one-factor scales. \textit{Task quality} splits into two subscales: \textit{goal clarity} and \textit{feasibility at home}. \textit{Authenticity} splits into two subscales in line with Ref.~\cite{Klein.2016}: \textit{reference to reality} and \textit{disciplinary authenticity}, whereby item Auth3 with equal load on both factors was assigned to the factor suggested by Ref.~\cite{Klein.2016}. \textit{Experience during the tasks} is divided into three subscales: \textit{experience of competence}, \textit{curiosity/interest (short)}, and \textit{autonomy/creativity (short)}; items EdT3 and EdT4 were excluded due to differing interpretations of their orientation. \textit{Perceived educational effectiveness} splits into two subscales: \textit{perceived cognitive effectiveness} and \textit{perceived affective effectiveness}. \textit{Autonomy} forms a satisfactory scale after excluding item Aut2, where not all participants recognized its inversion.

The linguistic and semantic similarity of some subscales built (e.g., \textit{autonomy} and \textit{autonomy/creativity (short)}) reflects the exploratory use and analysis of several instruments from different origins in this study. We report an analysis of all these subscales showcasing how the tasks can be coherently evaluated using several scales. It may also function as a resource for instructors and researchers who want to use and adapt these scales to their own needs.

\subsection{Data processing and method of data analysis}

\subsubsection{Analysis of weekly surveys (RQ1)}\label{MethodQuanAnalysis1a}

In the weekly surveys, 0\% to 12\% of students ($M=7\%$, $\mathrm{SD}=4\%$) each indicated no contribution to most group discussions and the task in general. Their responses are excluded from further analysis of respective surveys. Reported \textit{time spent on task} was winsorized to exclude unrealistic values below 5~min and above 300~min and quartiles are considered. For all other (sub-)scales or items in the weekly surveys, means and standard deviations are calculated. To go beyond the perceptions of the individual tasks, i.e. to compare the perceptions of learning with the Exp- and Pro-tasks similarly to the affective responses in the comparative surveys, the mean and standard deviation over the means (for times: medians) of all nine Exp- and three Pro-tasks are determined each and a Mann-Whitney-U test is performed. Responses to the three open feedback questions are analyzed separately for Exp- and Pro-tasks but combined across all surveys for each task type, regardless of multiple responses per person. Unrelated responses (e.g., positive feedback to what was disliked, or responses like "none") are excluded; the remaining are grouped into common themes. If a response has multiple aspects related to different themes, it is split into substatements and assigned accordingly, but not more than once to the same theme. Relative frequencies identify the most prominent themes.

\subsubsection{Analysis of comparative surveys (mostly RQ2)}\label{MethodQuanAnalysis2}

For the comparative surveys, responses to the (sub-)scales for the three different task types are tested for normal distribution with Shapiro-Wilk tests. As several scales are not normally distributed, analysis is performed nonparametrically. Bonferroni-corrected Wilcoxon tests applied to the scales used in both comparative surveys for the Exp-tasks show significant difference only for \textit{disciplinary authenticity} ($Z=3.14$, $p=0.001$, $p_B=0.008$, $N=34$). This was rated as $M=4.25$ ($\mathrm{SD}=0.74$) in Com~1 and as $M=3.77$ ($\mathrm{SD}=0.90$) in Com~2. It may be attributed to the two different comparisons required in the two surveys (Exp vs Pro and Exp vs Rec) and the timing of Com~2 close to the written exam including Rec-tasks only, potentially leading to a more limited perspective on disciplinary authentic tasks at that time.

As there were no differences in other variables, responses to the scales for the Exp-tasks used in both comparative surveys are combined to allow for a more comprehensive comparison and enhance data quality. If a student participated in only one comparative survey, the single scale score is used; otherwise, the average of the two scale scores is calculated.

The three task types are compared using Bonferroni-corrected Wilcoxon and Friedman tests. For significant Friedman tests, Bonferroni-corrected post-hoc analysis is performed to examine pairwise significant differences. In all tests reported here and above, cases are excluded pairwise to maximize the number of responses for each test.

\subsubsection{Analysis of the influence of predictors (RQ3)}\label{MethodQuanAnalysis3}

To analyze the influence of surveyed predictors, we reduced the number of dependent metrically scaled variables to broader parent variables through factor analysis. To maximize comparisons between the three task types, this analysis was conducted for three different groups of variables based on which were measured for the same task types (cf. data structure in Table~\ref{Tab:Variables} and Table~\ref{tab:factoranalysisRG3} in the Appendix). The factor analysis utilized z-standardized responses from Exp-tasks. Multiple responses per participant for \textit{goal clarity}, \textit{feasibility at home}, and \textit{overall task rating} were averaged to a single score for both the Exp- and Pro-tasks.
The analyses (cf. Table~\ref{tab:factoranalysisRG3} in the Appendix) indicate three parent variables. \textit{(P1) Task impression} (average of \textit{overall task rating}, \textit{goal clarity}, \textit{feasibility at home}, and \textit{use of technologies}) reflects students' overall perceptions of learning with the Exp- and Pro-tasks. \textit{(P2) Task value} (average of \textit{linking to the lecture}, \textit{perceived cognitive effectiveness}, and \textit{perceived affective effectiveness}) captures students' perceptions of the added value or benefit of Exp- and Rec-tasks. \textit{(P3) Affective impact} (average of \textit{curiosity}, \textit{interest}, \textit{reference to reality}, \textit{disciplinary authenticity}, \textit{curiosity/interest (short)}, and \textit{autonomy/creativity (short)}) generalizes the affective responses to the tasks.

Data on all predictors and parent variables for all task types were available for only 19 students, which is below the necessary sample size for MANCOVAs. Therefore, differences in the factors \textit{gender}, \textit{semester}, and \textit{study program} were explored using Bonferroni-corrected Mann-Whitney-U and Kruskal-Wallis tests. Spearman's $\rho$ correlations were calculated between the seven other predictors treated as covariates and the parent variables for the different task types, as well as between the parent variables for each task type, since they can also serve as covariates. In these analyses, for covariates measured in multiple surveys, the average of all responses per person was used. The open-text responses about \textit{prior experience with physics and smartphone experiments} were categorized on an ordinal scale as outlined in Table~\ref{tab:priorexperience} in the Appendix. Besides prior experience, covariates were z-standardized.

\section{Results}\label{results}

\subsection{Students' perceptions of learning with the tasks (RQ1)}

%\subsubsection{Closed task ratings}

Figure~\ref{fig:OverallImpression} and Fig.~\ref{fig:OverallImpressionLong} in the Appendix show the students' \textit{overall task ratings} of the Exp- and Pro-tasks on the artificial scale from 1 (worst) to 10 (best). Ratings range from $M=5.6$ ($\mathrm{SD}=2.4$) for Pro~1 to $M=7.1$ ($\mathrm{SD}=2.10$) for Pro~2. Mann-Whitney-U test shows no significant difference between the average ratings of the nine Exp-tasks and three Pro-tasks ($U=9.00,\,Z=0.83,\,p=0.46,\,r=0.24$).

\begin{figure}[htb]
\flushleft 
\begin{tikzpicture}
\begin{axis}[width=.73\columnwidth, height=2.8cm, xbar=0pt,
  ymax=.5,  
  xmin=1, xmax=10,
  xtick={1,2,3,4,5,6,7,8,9,10},
  ymin =0.1,
  xlabel={Mean \& standard deviation},
extra x ticks={1,10},
extra x tick labels={worst, best},
extra x tick style={grid=none, tick style={draw=none}, tick label style={xshift=-5pt, yshift=-8pt}},
  ytick = {0.2,0.4,.6,.8,1.,1.2,1.4,1.6,1.8,2.,2.2,2.4,2.6,2.8},
  yticklabel style={text width=.4\columnwidth,align=right, },
  yticklabels={Mean Pro 1-3, Mean Exp 1-9},
   ytick pos=left,
xtick pos=left,
 legend columns=1, legend cell align = left,legend style = {draw = none},
 legend style = {at ={(-0.15,1)}, anchor = south west},
 bar width = 7pt,
  ]
\addplot+[purple!60!,area legend,
    draw=black, error bars/.cd, x dir=both, x explicit, error mark options={black,mark size=2pt,line width=.7pt,rotate=90
     },  error bar style={line width=.7pt}
      ] 
		coordinates{
(	6.13, .2)+-(0.82, .2)
(	6.45, .4)+-(0.19, .4)
}; \label{overall}

\end{axis}
\end{tikzpicture}
\vspace{-0.7cm}\caption{Students' \textit{overall task rating} averaged across the mean ratings of the nine experimental and three programming tasks.% Values for all tasks are shown in Fig.~\ref{fig:OverallImpressionLong} in the Appendix.}% The dashed line indicates the scale middle.
}
\label{fig:OverallImpression}
\end{figure}
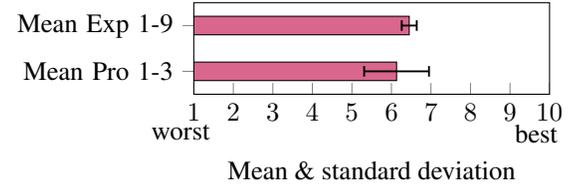

Figure~\ref{fig:Time} and Fig.~\ref{fig:TimeLong} in the Appendix show considerable variation in students' self-reported \textit{time spent on tasks}. The shortest median time was 20~min (Exp~2), while the longest was 90~min (Pro~1). The lower 25\% quartile was 15~min for Exp~1, Exp~2, Exp~4, and Pro~2, while the upper 75\% quartile was 120~min for Pro~1, indicating that most students spent no more than two hours on any task, typically much less. Mann-Whitney-U test revealed no significant difference regarding the average \textit{time spent on task} between the nine Exp-tasks and three Pro-tasks ($U=9.00,\,Z=0.85,\,p=0.43,\,r=0.25$).

\begin{figure}[htb]
\flushleft 
\begin{tikzpicture}
\begin{axis}[width=.73\columnwidth, height=2.8cm, xbar=0pt,
  ymax=.5,  
  xmin=0, xmax=90,
  xtick={0,30,60,90},
  ymin =0.1,
  xlabel={Mean \& standard deviation (in min)},
  ytick = {0.2,0.4,.6,.8,1.,1.2,1.4,1.6,1.8,2.,2.2,2.4,2.6,2.8},
  yticklabel style={text width=.4\columnwidth,align=right, },
  yticklabels={Mean Pro 1-3, Mean Exp 1-9},
   ytick pos=left,
xtick pos=left,
 legend columns=1, legend cell align = left,legend style = {draw = none},
 legend style = {at ={(-0.15,1)}, anchor = south west},
 bar width = 7pt,
  ]

\addplot+[purple!60!,area legend,
    draw=black, error bars/.cd, x dir=both, x explicit, error mark options={black,mark size=2pt,line width=.7pt,rotate=90
     },  error bar style={line width=.7pt}
      ] 
		coordinates{
(	55.00, .2)+-(19.13, .2)
(	38.33, .4)+-(12.25, .4)
}; \label{time}

\end{axis}
\end{tikzpicture}
\vspace{-0.7cm}\caption{Students' self-reported time spent, on average, on the nine experimental and three programming tasks.}% Values for all tasks are shown in Fig.~\ref{fig:TimeLong} in the Appendix.}
\label{fig:Time}
\end{figure}
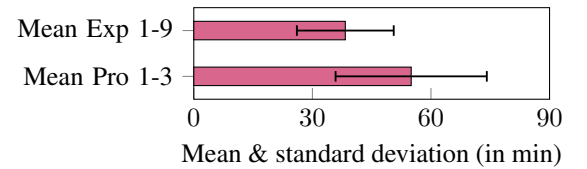

Figure~\ref{fig:Adequacyscale} and Fig.~\ref{fig:AdequacyscaleLong} in the Appendix show students' perceptions of \textit{goal clarity} and \textit{feasibility at home} of the Exp- and Pro-tasks. With all scores above 3.5, indicating an agreement, task instructions were generally clear and suitable for homework. Mann-Whitney-U test revealed no significant difference between the average scores of the nine Exp- and three Pro-tasks for \textit{feasibility at home} ($U=6.00,\,Z=1.39,\,p=0.21,\,r=0.40$) but for \textit{goal clarity} ($U=2.00,\,Z=2.13,\,p=0.04,\,r=0.61$).

\begin{figure}[htb]
\flushleft 
\begin{tikzpicture}
\begin{axis}[width=.73\columnwidth, height=3.2cm, xbar=0pt,
  ymax=.5,  
  xmin=1, xmax=5,
  xtick={1,2,3,4,5},
  ymin =0.1,
  xlabel={Mean \& standard deviation},
extra x ticks={1,5},
extra x tick labels={strongly disagree, strongly agree},
extra x tick style={grid=none, tick style={draw=none}, tick label style={xshift=-25pt, yshift=-8pt}},
  ytick = {0.2,0.4,.6,.8,1.,1.2,1.4,1.6,1.8,2.,2.2,2.4,2.6,2.8},
  yticklabel style={text width=.4\columnwidth,align=right, },
  yticklabels={Mean Pro 1-3, Mean Exp 1-9},
   ytick pos=left,
xtick pos=left,
 legend columns=1, legend cell align = left,legend style = {draw = none},
 legend style = {at ={(-0.19,1)}, anchor = south west},
 bar width = 5pt,
  ]
\addlegendimage{empty legend}
\addlegendentry{\scalebox{1}[1]{\ref{Clarity}} Goal clarity}
\addlegendimage{empty legend}
\addlegendentry{\scalebox{1}[1]{\ref{Feasibility}} Feasibility at home}

\addplot+[purple!30!,area legend,
    draw=black, error bars/.cd, x dir=both, x explicit, error mark options={black,mark size=2pt,line width=.7pt,rotate=90
     },  error bar style={line width=.7pt}
      ] 
		coordinates{
(	3.84, .2)+-(0.22, .2)
(	4.02, .4)+-(0.18, .4)
}; \label{Feasibility}

\addplot+[purple!60!,area legend,
    draw=black, error bars/.cd, x dir=both, x explicit, error mark options={black,mark size=2pt,line width=.7pt,rotate=90
     },  error bar style={line width=.7pt}
      ] 
		coordinates{
(	3.74, .2)+-(0.24, .2)
(	4.03, .4)+-(0.08, .4)
}; \label{Clarity}

\end{axis}
\end{tikzpicture}
\vspace{-0.7cm}\caption{Average students' perceptions of \textit{goal clarity} and \textit{feasibility at home} for the nine experimental and three programming tasks. %The dashed line indicates the scale middle. 
%The values for all tasks are shown in Fig.~\ref{fig:AdequacyscaleLong}.
}
\label{fig:Adequacyscale}
\end{figure}
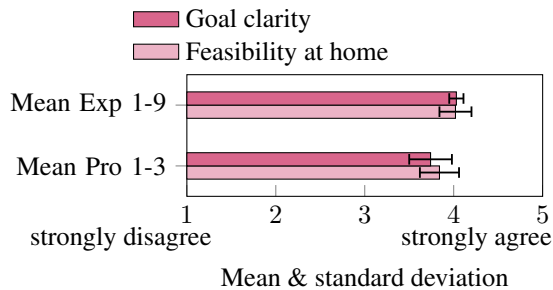

Figure~\ref{fig:difficultyInstructions} and Fig.~\ref{fig:difficultyInstructionslong} in the Appendix show the percentage of students who found the tasks too easy or too difficult and the task instructions insufficient. More than 20\% of the students found the tasks Exp~1, Exp~2, Exp~7, and Pro~2 too easy. The only task perceived by more than 20\% of the students as too difficult (43\%) and coming along with insufficient instructions (25\%) was Pro~1. For most students, all other tasks were adequately challenging and had sufficient task instructions.

\begin{figure}[htb]
\flushleft 
\begin{tikzpicture}
\begin{axis}[width=.73\columnwidth, height=3.5cm, xbar=0pt,
  ymax=.5,  
  xmin=0, xmax=45,
  xtick={0,10,20,30,40},
  ymin =0.1,
  xlabel={Percentage (\%)},
  ytick = {0.2,0.4,.6,.8,1.,1.2,1.4,1.6,1.8,2.,2.2,2.4,2.6,2.8},
  yticklabel style={text width=.4\columnwidth,align=right, },
  yticklabels={Mean Pro 1-3, Mean Exp 1-9},
   ytick pos=left,
xtick pos=left,
 legend columns=1, legend cell align = left,legend style = {draw = none},
 legend style = {at ={(-0.19,1)}, anchor = south west},
 bar width = 5pt,
  ]
\addlegendimage{empty legend}
\addlegendentry{\scalebox{1}[1]{\ref{Easy}} Task too easy}
\addlegendimage{empty legend}
\addlegendentry{\scalebox{1}[1]{\ref{Difficult}} Task too difficult}
\addlegendimage{empty legend}
\addlegendentry{\scalebox{1}[1]{\ref{Insufficient}} Task instructions insufficient}

\addplot+[yellow!60!,area legend,
    draw=black, error bars/.cd, x dir=both, x explicit, error mark options={black,mark size=2pt,line width=.7pt,rotate=90
     },  error bar style={line width=.7pt}
      ] 
		coordinates{
(	16.96, .2)+-(8.35, .2)
(	7.64, .4)+-(4.14, .4)
}; \label{Insufficient}

\addplot+[red!60!,area legend,
    draw=black, error bars/.cd, x dir=both, x explicit, error mark options={black,mark size=2pt,line width=.7pt,rotate=90
     },  error bar style={line width=.7pt}
      ] 
		coordinates{
(	22.60, .2)+-(17.51, .2)
(	3.60, .4)+-(5.31, .4)
}; \label{Difficult}

\addplot+[green!60!,area legend,
    draw=black, error bars/.cd, x dir=both, x explicit, error mark options={black,mark size=2pt,line width=.7pt,rotate=90
     },  error bar style={line width=.7pt}
      ] 
		coordinates{
(	8.73, .2)+-(11.06, .2)
(	17.22, .4)+-(9.88, .4)
}; \label{Easy}

\end{axis}
\end{tikzpicture}
\vspace{-0.7cm}\caption{Average percentage of students (mean and standard deviation) who perceived the nine experimental and three programming tasks as too easy or difficult and the task instructions as insufficient.% The values for all tasks are shown in Fig.~\ref{fig:difficultyInstructionslong} in the Appendix.
}
\label{fig:difficultyInstructions}
\end{figure}
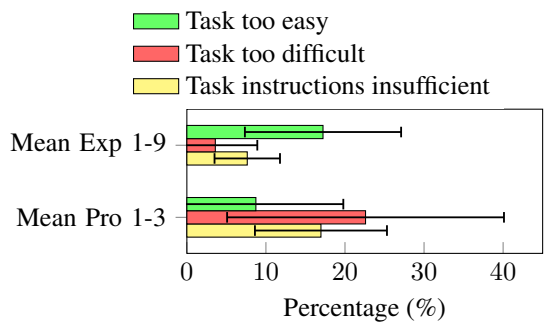

%\subsubsection{Use of technologies}

$N=87$ students rated the use of technologies, i.e. the helpfulness and usefulness of smartphones in the Exp-tasks, as $M=3.71$ ($\mathrm{SD}=0.63$) on a scale of 1 (\textit{strongly disagree}) to 5 (\textit{strongly agree}). Regarding Python in the Pro-tasks, the rating is $M=3.54$ ($\mathrm{SD}=0.79$). Wilcoxon test shows no significant difference ($Z=1.18$, $p=0.24$, $r=0.13$).

%\subsubsection{Open task feedback}

\begin{table*}[htb]
\caption{Summary of open task feedback from the twelve weekly surveys regarding what students (multiple responses per student) liked and disliked about the experimental and programming tasks. In each row, semantically related positive and negative aspects are compared.}
\begin{ruledtabular}
\footnotesize
\begin{tabular}{p{.395\textwidth}p{.04\textwidth}p{.04\textwidth}p{.395\textwidth}p{.04\textwidth}p{.04\textwidth}}
What the students liked about the tasks:&Exp&Pro&What the students disliked about the tasks:&Exp&Pro\\\hline
Absolute number of responses&440&112&Absolute number of responses&396&95\\
of which contain positive statements&425&93&of which contain negative statements&339&66\\\hline
Number of categorized (sub-)statements&557&110&Number of categorized (sub-)statements&380&72\\\hline
\textbf{Frequency of each topic in the positive (sub-)statements}&in \%&in \% &\textbf{Frequency of each topic in the negative (sub-)statements}&in \%&in \%\\
Complexity adequate, (time) expenditure adequate/low&3&7&Complexity and/or (time) expenditure too high&4&29\\
\raggedright\hangindent=.3cm Task (in general) simple, easy to understand&12&17&Task (in general) too easy/undemanding&4&1\\
\raggedright\hangindent=.3cm Task instructions good, clear, easy to understand, helpful&4&19&Task instructions difficult to understand, insufficient&12&28\\
\raggedright\hangindent=.3cm Task (elements) interesting, new, relevant, fun&8&19&Task boring, not interesting, repetitive, nothing new&10&0\\
\raggedright\hangindent=.3cm Hands-on, practical, activating and independent experimentation that makes fun&10&N/A&\raggedright\hangindent=.3cm Insufficient circumstances for conduction, e.g., regarding availability of equipment, local conditions&13&N/A\\
Simple setup, not much or only simple equipment needed&9&N/A&Difficulties with or extent of building and adjusting the setup&6&N/A\\
\raggedright\hangindent=.3cm \textit{phyphox}, external sensors, and smartphones for experiments&6&N/A&\raggedright\hangindent=.3cm Technical difficulties or disliking \textit{phyphox}/external sensors&6&N/A\\
Simple/Quick conduction \& practicability of experiment/task&17&N/A&Difficulties during the conduction of the experiment&12&N/A\\
Data analysis/Programming&3&10&Disliking or difficulties with data analysis/programming&2&19\\
\raggedright\hangindent=.3cm Quality/Reliability/Comprehensibility of measuring process and results&4&N/A&\raggedright\hangindent=.3cm Lack of precision and measurement accuracy, high susceptibility to errors and uncertainties&12&N/A\\
Linking theory to practice, deepening (prior) knowledge&4&13&Missing prior knowledge/expertise&0&4\\
\raggedright\hangindent=.3cm Gain in knowledge/expertise/competency&2&4&\raggedright\hangindent=.3cm Added value/purpose unclear, poor effort-benefit ratio&6&0\\
Opportunities for being creative, making own decisions&7&2&Too much predetermined, insufficient flexibility&0&10\\
Group work&3&0&Difficulties regarding group size&2&0\\
Other comments&7&9&Other comments&8&8\\
\end{tabular}
\end{ruledtabular}
\label{Tab:positiveNegativeFeedback}
\end{table*}

Table~\ref{Tab:positiveNegativeFeedback} summarizes open-text responses in the weekly surveys on what students liked and disliked about the Exp- and Pro-tasks. About the Exp-tasks, they liked the simple conduction and practicability of the experiments (17\% of the 557~positive statements for Exp-tasks), the overall simplicity and ease of understanding the tasks (12\%), the activating hands-on experimentation (10\%), the simple setup (9\%), and that the tasks were interesting, relevant, and fun (8\%). Regarding the Pro-tasks, students similarly found the tasks interesting and fun (19\% of the 110~positive statements for Pro-tasks) as well as simple and easy to understand (17\%). Additionally, they liked the clear, easy-to-understand, and helpful task instructions and supportive materials (19\%), the linking of theory to practice (13\%), and the programming itself (10\%).

About the Exp-tasks, students disliked circumstances like availability of equipment (13\% of the 380~negative statements for Exp-tasks), the difficulty or insufficiency of task instructions (12\%), and the susceptible precision and accuracy of measurements (12\%). They perceived some tasks as not interesting (10\%) or had difficulties during the experiment conduction (12\%). The Pro-tasks (primarily task Pro~1) were perceived as too complex and time-demanding (29\% of the 72~negative statements for Pro-tasks) and lacking comprehensible or sufficient task instructions (28\%). Some students disliked the programming or had difficulties with it (19\%), or criticized the pre-written code as inflexible (10\%).

\begin{table}[htb]
\caption{Overview of open task feedback responses from the twelve weekly surveys regarding suggestions for improving the tasks. The responses (multiple responses per student) are thematically clustered and split for experimental and programming tasks.}
\begin{ruledtabular}
\footnotesize
\begin{tabular}{p{.83\columnwidth}p{.06\columnwidth}p{.06\columnwidth}}
What the students suggest as improvements for the tasks:&Exp&Pro\\\hline
Absolute number of responses&272&75\\
of which contain suggestions for improvement&164&47\\\hline
Number of categorized (sub-)statements&171&50\\\hline
\textbf{Frequency of each topic in the (sub-)statements}&in \%&in \%\\
\raggedright\hangindent=.3cm Improved equipment: availability, precision, suitability, ...&25&N/A\\
\raggedright\hangindent=.3cm More explanations: hints, specifications, control solutions, adapted to prior expertise&12&34\\
\raggedright\hangindent=.3cm Improved instructions: (goal) clarity, comprehensibility, ...&15&20\\
\raggedright\hangindent=.3cm Higher demand/depth of task: more variations, options to expend task, ...&8&10\\
Lower demand, in scope or complexity&6&18\\
\raggedright\hangindent=.3cm No/Less Exp-/Pro-tasks, partly in favor of more Rec-tasks&8&10\\
Specific hints about phyphox application&6&N/A\\
\raggedright\hangindent=.3cm Improved circumstances: availability of premises and time, conduction on-campus or at home, ...&7&0\\
\raggedright\hangindent=.3cm Specific improvements of experiment design or setup&4&N/A\\
More motivating, dynamic task&4&0\\
Other suggestions&4&8
\end{tabular}
\end{ruledtabular}
\label{Tab:suggestionsImprovements}
\end{table}

What students disliked about the tasks aligns with their suggestions for improvements (cf. Table~\ref{Tab:suggestionsImprovements}). Regarding the task instructions, they suggested enhanced precision, (goal) clarity, and overall comprehensibility. In response to the complexity of the Pro-tasks, they asked for lower demands and especially more explanations like hints, specifications, or control solutions. For the Exp-tasks, suggestions are about the improvement of the availability, suitability, and precision of the equipment. Some students requested more demanding or in-depth Exp- and Pro-tasks, while others would reduce or even eliminate these tasks, partly in favor of more Rec-tasks.

\subsection{Affective responses (RQ2)}

In the following, variables are grouped in diagrams according to rating scales and task types they pertain to. Bonferroni correction is applied for each diagram or post-hoc analysis.

\begin{figure}[htb]
\flushleft
\begin{tikzpicture}
\begin{axis}[width=.73\columnwidth, height=6.3cm, xbar=0pt,
  ymax=0.9,
  xmin=1, xmax=6,
  xtick={1,2,3,4,5,6},
  ymin =0.1,
  xlabel={Mean \& standard deviation},
  extra x ticks={1,6},
  extra x tick labels={low, high},
  extra x tick style={grid=none, tick style={draw=none}, tick label style={xshift=-7pt, yshift=-8pt}},
  ytick = {0.2,0.4,.6,.8},
  yticklabel style={text width=.4\columnwidth,align=right,},
  yticklabels={Disciplinary authenticity, Reference to reality, Interest, Curiosity},
  ytick pos=left,
  xtick pos=left,
  legend columns=1, legend cell align = left,legend style = {draw = none},
  legend style = {at ={(-0.19,1)}, anchor = south west},
  bar width = 7pt,
  nodes near coords,
  %nodes near coords align={left},
  every node near coord/.append style={font=\scriptsize, text=white, anchor=east, xshift=-30pt}, point meta=explicit symbolic,
  ]
\addlegendimage{empty legend}
\addlegendentry{\scalebox{1}[1]{\ref{Exp}} Experimental tasks (Exp)}
\addlegendimage{empty legend}
\addlegendentry{\scalebox{1}[1]{\ref{Pro}} Programming tasks (Pro)}
\addlegendimage{empty legend}
\addlegendentry{\scalebox{1}[1]{\ref{Rec}} Standard recitation tasks (Rec)}

\addplot+[blue!60!,area legend,
    draw=black, error bars/.cd, x dir=both, x explicit, error mark options={black,mark size=2pt,line width=.7pt,rotate=90
     }, error bar style={line width=.7pt},
      ] 
		coordinates{
(4.46, .2)+-(0.85,.2) [N=59]
(4.06, .4)+-(0.97, .4) [N=59]
(4.51, .6)+-(0.67, .6) [N=70]
(4.33, .8)+-(0.86, .8) [N=72]
}; \label{Rec}

\addplot+[gray,area legend,
    draw=black, error bars/.cd, x dir=both, x explicit, error mark options={black,mark size=2pt,line width=.7pt,rotate=90
     },  error bar style={line width=.7pt},
      ] 
		coordinates{
(3.65, .2)+-(1.05,.2) [N=92]
(2.96, .4)+-(0.92, .4) [N=92]
(3.15, .6)+-(1.08, .6) [N=101]
(3.15, .8)+-(1.08, .8) [N=105]
}; \label{Pro}

\addplot+[orange,area legend,
    draw=black, error bars/.cd, x dir=both, x explicit, error mark options={black,mark size=2pt,line width=.7pt,rotate=90
     },  error bar style={line width=.7pt},
      ] 
		coordinates{
(4.01, .2)+-(0.93,.2) [N=116]
(4.06, .4)+-(0.88, .4) [N=117]
(3.59, .6)+-(0.93, .6) [N=129]
(3.63, .8)+-(0.92, .8) [N=132]
}; \label{Exp}

\end{axis}
\end{tikzpicture}
\vspace{-0.7cm}\caption{Comparison between the three task types regarding the caused \textit{curiosity} and \textit{interest} as well as the perceived \textit{authenticity} (\textit{reference to reality} and \textit{disciplinary authenticity}).}% The dashed line indicates the scale middle.}
\label{fig:CurIntAuth}
\end{figure}
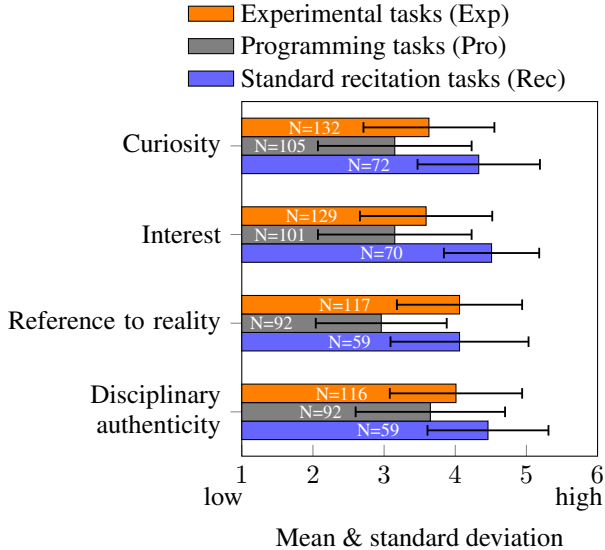

Figure~\ref{fig:CurIntAuth} shows the \textit{curiosity}, \textit{interest}, and \textit{authenticity} elicited by the three different task types as perceived by the students. Bonferroni-corrected Friedman tests show significant differences for all four (sub-)scales (cf. Table~\ref{Tab:FriedmanAffec} in the Appendix). Post-hoc analyses, Bonferroni-corrected in each case, show that \textit{curiosity} is significantly higher for the Rec-tasks than for the Exp- ($p<0.001$, $p_B=0.002$) and Pro-tasks ($p<0.001$, $p_B<0.001$). Interest is perceived as significantly higher for the Rec-tasks than the Exp-tasks ($p<0.001$, $p_B<0.001$) and Pro-tasks ($p<0.001$, $p_B<0.001$). \textit{Reference to reality} is significantly higher for the Rec-tasks ($p<0.001$, $p_B<0.001$) and the Exp-tasks ($p<0.001$, $p_B<0.001$) than the Pro-tasks. \textit{Disciplinary authenticity} is significantly higher for the Rec-tasks than for the Pro-tasks ($p=0.001$, $p_B=0.004$).

\begin{figure}[htb]
\flushleft 
\begin{tikzpicture}
\begin{axis}[width=.73\columnwidth, height=5cm, xbar=0pt,
  ymax=0.7,  
  xmin=1, xmax=5,
  xtick={1,2,3,4,5},
  ymin =0.1,
  xlabel={Mean \& standard deviation},
extra x ticks={1,5},
extra x tick labels={strongly disagree, strongly agree},
extra x tick style={grid=none, tick style={draw=none}, tick label style={xshift=-25pt, yshift=-8pt}},
  ytick = {0.2,0.4,.6},
  yticklabel style={text width=.4\columnwidth,align=right, },
  yticklabels={Autonomy/Creativity (short), Curiosity/Interest (short), Experience of competence},
   ytick pos=left,
xtick pos=left,
 legend columns=1, legend cell align = left,legend style = {draw = none},
 legend style = {at ={(-0.19,1)}, anchor = south west},
 bar width = 7pt,
   nodes near coords,
  %nodes near coords align={left},
  every node near coord/.append style={font=\scriptsize, text=white, anchor=east, xshift=-32pt}, point meta=explicit symbolic,
  ]
\addlegendimage{empty legend}
\addlegendentry{\scalebox{1}[1]{\ref{ExpEmo}} Experimental tasks (Exp)}
\addlegendimage{empty legend}
\addlegendentry{\scalebox{1}[1]{\ref{ProEmo}} Programming tasks (Pro)}
\addlegendimage{empty legend}
\addlegendentry{\scalebox{1}[1]{\ref{RecEmo}} Standard recitation tasks (Rec)}

\addplot+[blue!60!,area legend,
    draw=black, error bars/.cd, x dir=both, x explicit, error mark options={black,mark size=2pt,line width=.7pt,rotate=90
     },  error bar style={line width=.7pt}
      ] 
		coordinates{
(	3.26, .2)+-(0.93,.2) [N=51]
(	3.77, .4)+-(0.65, .4) [N=51]
(   3.22, .6)+-(0.71, .6) [N=51]
}; \label{RecEmo}

\addplot+[gray,area legend,
    draw=black, error bars/.cd, x dir=both, x explicit, error mark options={black,mark size=2pt,line width=.7pt,rotate=90
     },  error bar style={line width=.7pt}
      ] 
		coordinates{
(	2.99, .2)+-(0.98,.2) [N=98]
(   3.02, .4)+-(0.85, .4) [N=98]
(   2.80, .6)+-(0.98, .6) [N=98]
}; \label{ProEmo}

\addplot+[orange,area legend,
    draw=black, error bars/.cd, x dir=both, x explicit, error mark options={black,mark size=2pt,line width=.7pt,rotate=90
     },  error bar style={line width=.7pt}
      ] 
		coordinates{
(	3.38, .2)+-(0.83,.2) [N=118]
(	3.40, .4)+-(0.74, .4) [N=118]
(   3.39, .6)+-(0.83, .6) [N=118]
}; \label{ExpEmo}

\end{axis}
\end{tikzpicture}
\vspace{-0.7cm}\caption{Comparison between the three task types regarding the students' \textit{experience during the tasks}.}% The dashed line indicates the scale middle.}
\label{fig:emotions}
\end{figure}
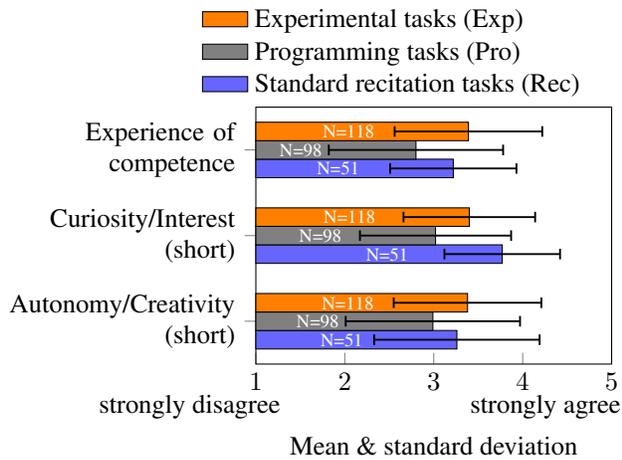

Figure~\ref{fig:emotions} compares the three task types across the subscales \textit{experience of competence}, \textit{curiosity/interest (short)}, and \textit{autonomy/creativity (short)}. Bonferroni-corrected Friedman tests (cf. Table~\ref{Tab:FriedmanAffec} in the Appendix) show significant differences for \textit{experience of competence} and \textit{curiosity/interest (short)} but not for \textit{autonomy/creativity (short)}. Bonferroni-corrected post-hoc analyses show higher \textit{experience of competence} during the Exp-tasks than during the Pro-tasks ($p<0.001$, $p_B=0.001$) and Rec-tasks ($p=0.003$, $p_B=0.009$). \textit{Curiosity/Interest (short)} was lower for the Pro-tasks than the Rec-tasks ($p=0.008$, $p_B=0.023$).

\begin{figure}[htb]
\flushleft 
\begin{tikzpicture}
\begin{axis}[width=.73\columnwidth, height=3.4cm, xbar=0pt,
  ymax=0.5,  
  xmin=1, xmax=6,
  xtick={1,2,3,4,5,6},
  ymin =0.1,
  xlabel={Mean \& standard deviation},
extra x ticks={1,6},
extra x tick labels={low, high},
extra x tick style={grid=none, tick style={draw=none}, tick label style={xshift=-7pt, yshift=-8pt}},
  ytick = {0.2,0.4,.6,.8},
  yticklabel style={text width=.4\columnwidth,align=right, },
  yticklabels={Linking to the lecture, Autonomy, Perceived affective effectiveness (N=74), Perceived cognitive effectiveness (N=74)},
   ytick pos=left,
xtick pos=left,
 legend columns=1, legend cell align = left,legend style = {draw = none},
 legend style = {at ={(-0.19,1)}, anchor = south west},
 bar width = 7pt,
    nodes near coords,
  %nodes near coords align={left},
  every node near coord/.append style={font=\scriptsize, text=white, anchor=east, xshift=-32pt}, point meta=explicit symbolic,
  ]
\addlegendimage{empty legend}
\addlegendentry{\scalebox{1}[1]{\ref{ExpAut}} Experimental tasks (Exp)}
\addlegendimage{empty legend}
\addlegendentry{\scalebox{1}[1]{\ref{RecAut}} Standard recitation tasks (Rec)}

\addplot+[blue!60!,area legend,
    draw=black, error bars/.cd, x dir=both, x explicit, error mark options={black,mark size=2pt,line width=.7pt,rotate=90
     },  error bar style={line width=.7pt}
      ] 
		coordinates{
(	5.15, .2)+-(0.66,.2) [N=66]%Linking
(	4.05, .4)+-(1.08, .4) [N=74]%Aut
}; \label{RecAut}

\addplot+[orange,area legend,
    draw=black, error bars/.cd, x dir=both, x explicit, error mark options={black,mark size=2pt,line width=.7pt,rotate=90
     },  error bar style={line width=.7pt}
      ] 
		coordinates{
(	4.43, .2)+-(0.88,.2) [N=66]%Linking
(	4.15, .4)+-(1.04, .4) [N=74]%Aut
}; \label{ExpAut}

\end{axis}
\end{tikzpicture}
\vspace{-0.7cm}\caption{Comparison between experimental and standard recitation tasks regarding the perceived \textit{autonomy} and \textit{linking to the lecture}.}% The dashed line indicates the scale middle.}
\label{fig:Aut}
\end{figure}
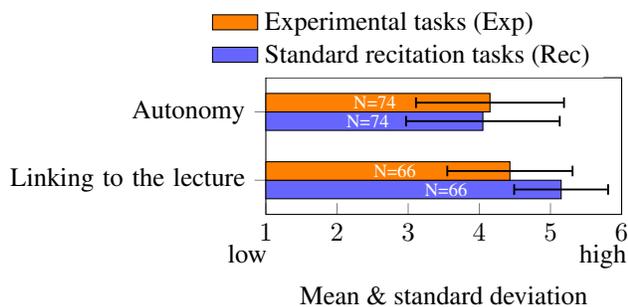

Figures~\ref{fig:Eff} and \ref{fig:Aut} show students' perceptions of the Exp- and Rec-tasks in terms of perceived educational effectiveness, \textit{autonomy}, and \textit{linking to the lecture}. Wilcoxon tests revealed no significant differences for \textit{perceived cognitive effectiveness} ($Z=1.41$, $p=0.16$, $p_B=0.32$, $N=74$) and \textit{autonomy} ($Z=0.77$, $p=0.44$, $p_B=0.89$, $N=74$), but Bonferroni-corrected significant differences for \textit{perceived affective effectiveness} ($Z=4.40$, $p<0.001$, $p_B<0.001$, $N=74$) and \textit{linking to the lecture} ($Z=5.86$, $p<0.001$, $p_B<0.001$, $N=66$), both higher rated for Rec-tasks.

\begin{figure}[htb]
\flushleft 
\begin{tikzpicture}
\begin{axis}[width=.73\columnwidth, height=3.5cm, xbar=0pt,
  ymax=0.9,  
  xmin=1, xmax=5,
  xtick={1,2,3,4,5},
  ymin =0.5,
  xlabel={Mean \& standard deviation},
extra x ticks={1,5},
extra x tick labels={strongly disagree, strongly agree},
extra x tick style={grid=none, tick style={draw=none}, tick label style={xshift=-25pt, yshift=-8pt}},
  ytick = {0.2,0.4,.6,.8},
  yticklabel style={text width=.4\columnwidth,align=right, },
  yticklabels={Linking to the lecture (N=66), Autonomy (N=74), Perceived affective effectiveness, Perceived cognitive effectiveness},
   ytick pos=left,
xtick pos=left,
 legend columns=1, legend cell align = left,legend style = {draw = none},
 legend style = {at ={(-0.19,1)}, anchor = south west},
 bar width = 7pt,
    nodes near coords,
  %nodes near coords align={left},
  every node near coord/.append style={font=\scriptsize, text=white, anchor=east, xshift=-32pt}, point meta=explicit symbolic,
  ]
\addlegendimage{empty legend}
\addlegendentry{\scalebox{1}[1]{\ref{ExpEff}} Experimental tasks (Exp)}
\addlegendimage{empty legend}
\addlegendentry{\scalebox{1}[1]{\ref{RecEff}} Standard recitation tasks (Rec)}

\addplot+[blue!60!,area legend,
    draw=black, error bars/.cd, x dir=both, x explicit, error mark options={black,mark size=2pt,line width=.7pt,rotate=90
     },  error bar style={line width=.7pt}
      ] 
		coordinates{
%(	5.15, .2)+-(0.66,.2)
%(	3.88, .4)+-(0.96, .4)
%(   4.58, .6)+-(0.79, .6)%with transformation &transformation to a scale from 1-6: -1, /4, *5, +1; for error bars: /4, *5
%(   4.80, .8)+-(0.79, .8)%with transformation
(   3.86, .6)+-(0.63, .6) [N=74]%without transformation
(   4.04, .8)+-(0.63, .8) [N=74]%without transformation
}; \label{RecEff}

\addplot+[orange,area legend,
    draw=black, error bars/.cd, x dir=both, x explicit, error mark options={black,mark size=2pt,line width=.7pt,rotate=90
     },  error bar style={line width=.7pt}
      ] 
		coordinates{
%(	4.43, .2)+-(0.88,.2)
%(	3.91, .4)+-(0.85, .4)
%(   3.99, .6)+-(1.09, .6)%with transformation
%(   4.69, .8)+-(0.80, .8)%with transformation
(   3.39, .6)+-(0.87, .6) [N=74]%without transformation
(   3.95, .8)+-(0.64, .8) [N=74]%without transformation
}; \label{ExpEff}

\end{axis}
\end{tikzpicture}
\vspace{-0.7cm}\caption{Comparison between the experimental and standard recitation tasks regarding the perceived educational effectiveness.}% The dashed line indicates the scale middle.}
\label{fig:Eff}
\end{figure}
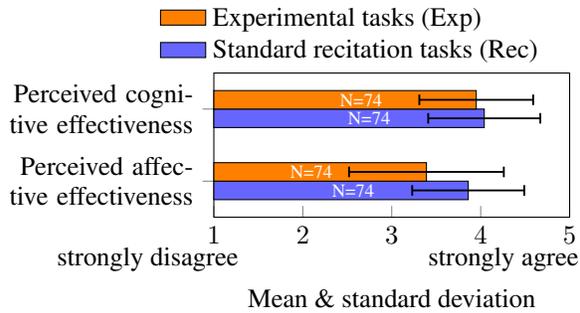

\subsection{Influence of predictors (RQ3)}

Influence of predictors was analyzed using the parent variables \textit{(P1) task impression} for the Exp- and Pro-tasks, \textit{(P2) task value} for the Exp- and Rec-tasks, and \textit{(P3) affective impact} for all three task types (cf. Sec.~\ref{MethodQuanAnalysis3}). For \textit{gender} and \textit{study program} (3-years Physics, 4-years Physics, and teacher training), Bonferroni-corrected Mann-Whitney-U and Kruskal-Wallis tests revealed no significant differences for any parent variable. For \textit{semester}, Mann-Whitney-U test was significant only for \textit{(P1) task impression} of the Pro-tasks ($U=20$, $Z=3.34$, $p<0.001$, $p_B=0.001$, $N=50$); first-semester students rated the \textit{(P1) task impression} higher ($M=0.12$, $\mathrm{SD}=0.58$, $N=44$) compared to higher-semester students ($M=-0.89$, $\mathrm{SD}=0.53$, $N=6$).

\begin{figure*}
    \centering
    \includegraphics[width=\textwidth]{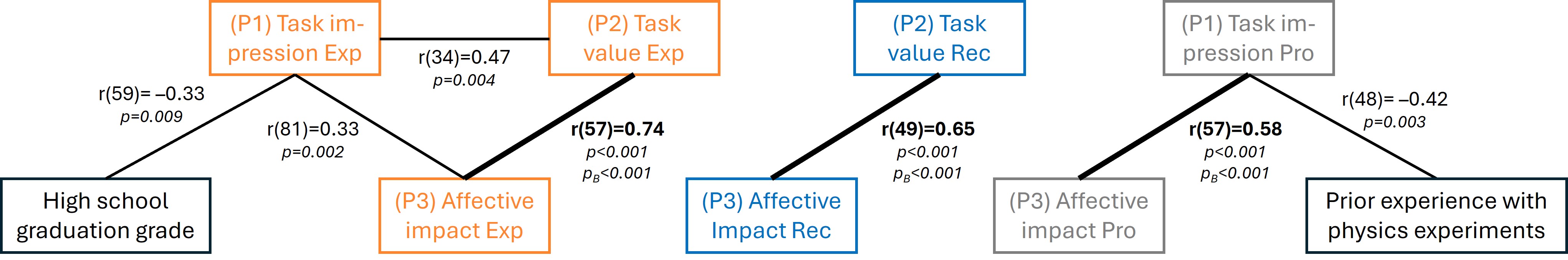}
    \vspace{-0.7cm}\caption{Spearman's $\rho$ correlations between the parent variables -- \textit{(P1) task impression}, \textit{(P2) task value}, and \textit{(P3) affective impact} for experimental, programming, and standard recitation tasks -- and covariates that become significant at the $p<0.01$ level before Bonferroni-correction; highlighted correlations are significant even after Bonferroni-correction for 55 examined correlations.}
    \label{fig:Correlations}
\end{figure*}

Spearman's $\rho$ correlations were computed between the seven predictors treated as covariates and the parent variables; additionally, five correlations between parent variables for the same task type were calculated. Out of a total of 55 correlations, only 7 were significant at the $p<0.01$ level without Bonferroni-correction, and just 3 at the $p<0.001$ level, even after Bonferroni-correction. Among these seven correlations (cf. Fig.~\ref{fig:Correlations}) are all five correlations between the parent variables. Additionally, \textit{high school graduation grade} (with lower numbers indicating better grades in Germany) showed an inverse correlation with \textit{(P2) task impression} of the Exp-tasks, and \textit{prior experience with physics experiments} correlates inversely with \textit{(P2) task impression} of the Pro-tasks.

\section{Discussion}

Table~\ref{Tab:overallcomparison} summarizes the statistical comparisons between the three task types. It shows that the Exp-tasks were perceived as equal or superior to the Pro-tasks in all aspects. Compared to the Rec-tasks, the Exp- and Pro-tasks were perceived as equal or inferior regarding all variables, except for \textit{experience of competence}, which was rated higher for the Exp-tasks than for the Rec-tasks. Overall, this indicates that students (slightly) preferred the Rec- to the Exp-tasks, and the Exp- to the Pro-tasks (Rec $\geq$ Exp $\geq$ Pro).

\begin{table}[htb]
\caption{Summary of the analysis in Sec.~\ref{results} comparing experimental, programming, and standard recitation tasks. $=$ indicates no significant difference; $>$ or $<$ indicate that the first task type was perceived significantly ($p_B < 0.05$) higher or lower than the second.}
\begin{ruledtabular}
\footnotesize
\begin{tabular}{p{.44\columnwidth} >{\centering}p{.155\columnwidth} >{\centering}p{.16\columnwidth} cp{.155\columnwidth}}
Variable&Exp vs Pro&Exp vs Rec&Pro vs Rec\\\hline
\multicolumn{3}{l}{\textbf{Students' perceptions of learning with the tasks (RQ1)}}&\\
Overall task rating&$=$&&\\
Goal clarity&$>$&&\\
Feasibility at home&$=$&&\\
Time spent on tasks&$=$&&\\
Use of technologies&$=$&&\\
\textbf{Affective responses (RQ2)}&&&\\
Curiosity&$=$&$<$&$<$\\
Interest&$=$&$<$&$<$\\
Reference to reality&$>$&$=$&$<$\\
Disciplinary authenticity&$=$&$=$&$<$\\
Experience of competence&$>$&$>$&$=$\\
Curiosity/Interest (short)&$=$&$=$&$<$\\
Autonomy/Creativity (short)&$=$&$=$&$=$\\
%\textbf{RG2b}&&&\\
Perceived cognitive effectiveness&&$=$&\\
Perceived affective effectiveness&&$<$&\\
Autonomy&&$=$&\\
Linking to the lecture&&$<$&\\
\end{tabular}
\end{ruledtabular}
\label{Tab:overallcomparison}
\end{table}

\subsection{Students' perceptions of learning with the tasks (RQ1)}

Analysis of students' perceptions of learning with the tasks reveals no larger differences between Exp- and Pro-tasks or between tasks of the same task type. \textit{Overall task rating}, \textit{goal clarity}, and \textit{feasibility at home} are generally assessed as medium to good across all tasks. The trend of increasing perceived \textit{task difficulty} over the semester aligns with the actual design of increasing task complexity culminating in the last two tasks Exp~8 and Exp~9 that combine experimentation and Python-aided data analysis. Students' positive ratings for the \textit{use of technologies} suggest that smartphones and Python were seen as useful and manageable digital tools for the tasks. These findings show that, although most Exp- and Pro-tasks were implemented into teaching for the first time, their design was already very suitable for the course and target group, often comparable to Rec-tasks, established and improved over years. 
The only task that exceeded the ability of a higher proportion of students was Pro~1, rated as too difficult by 43\% of students. This is also reflected in a lower \textit{overall task rating}, higher \textit{time spent on task}, and slightly lower \textit{goal clarity}.

Reported \textit{time spent on task} varies strongly among participants. This could be due to differences in how group members divided the workload, distorted time perceptions, socially desirable responses, or differences in whether students considered only the experiment or the entire task process when reporting times. The median time ranges from 20~min to 90~min, which is mostly in the expected range or below.

Open student feedback from the weekly surveys reveal a mix of positive recognition and contrasting perspectives regarding the Exp-tasks. On the one hand, students appreciated the intended positive effects of these tasks, such as the opportunities for hands-on experimentation using simple equipment, quick execution of physics experiments, fostering interest, and linking theory to practice. On the other hand, responses reveal contrasting perspectives on the tasks. For instance, some students appreciated the simplicity and quick execution of certain experiments, while others faced difficulties during their conduction. Similarly, some found the Pro-task instructions clear and comprehensible, while others criticized them as confusing or insufficient. This ambiguity reflects inherent differences in task design as well as varied perceptions and attitudes among students. It underscores the complexity of designing high-quality experimental and programming tasks that meet students' diverse needs and expectations.

\subsection{Affective responses (RQ2)}

\begin{figure*}[htb]
\flushleft
\begin{tikzpicture}
\begin{axis}[width=.86\textwidth, height=7.8cm, xbar=0pt,
  ymax=1.1,
  xmin=-1.8, xmax=1.8,  xtick={-1.8,-1.6,-1.4,-1.2,-1,-0.8,-0.6,-0.4,-0.2,0,0.2,0.4,0.6,0.8,1,1.2,1.4,1.6,1.8},
  ymin =0.1,
  xlabel={Effect size Cohen's $d_p$ with 95\% confidence interval},
  ytick = {0.2,0.4,.6,.8,1},
  yticklabel style={text width=.4\columnwidth,align=right,},
  yticklabels={Autonomy, Disciplinary authenticity, Reference to reality, Interest, Curiosity},
  ytick pos=left,
  xtick pos=left,
  legend columns=1, legend cell align = left,legend style = {draw = none},
  legend style = {at ={(-0.007,1)}, anchor = south west}, bar width = 7pt,
        point meta=explicit symbolic,
]
\addlegendimage{area legend, fill=pink!60!, draw=pink}
\addlegendentry{\,Experimental task (Exp) vs standard recitation tasks (Rec)}
\addlegendimage{area legend, fill=purple!60!, draw=purple}
\addlegendentry{\,Experimental tasks (Exp) vs comparative data for video analysis tasks \cite{Klein.2016}}
\addlegendimage{area legend, fill=cyan!50!, draw=cyan}
\addlegendentry{\,Experimental task (Exp) vs programming tasks (Pro)}
\addlegendimage{area legend, fill=blue!60!, draw=blue}
\addlegendentry{\,Standard recitation tasks (Rec) vs comparative data for standard recitation tasks \cite{Klein.2016}}

\draw[dashed] (230,0.1) -- (230,1000);
\draw[dashed] (260,0.1) -- (260,1000);
\draw[dashed] (130,0.1) -- (130,1000);
\draw[dashed] (100,0.1) -- (100,1000);
\draw[dashed] (200,0.1) -- (200,1000);
\draw[dashed] (160,0.1) -- (160,1000);

\addplot+[    fill=blue!60!, draw=blue, line width=1pt, error bars/.cd, 
    x dir=both, 
    x explicit,
    error bar style={line width=.7pt},
    error mark options={mark size=2pt,line width=.7pt,rotate=90, blue},
      ] 
		coordinates{
(0.05, .2)+-(0.385,.2)
(0.084, .4)+-(0.402, .4)
(0.184, .6)+-(0.402, .6)
(0.175, .8)+-(0.389, .8)
(0.48, 1)+-(0.392, 1)
}; \label{RecRecd}

\addplot+[fill=cyan!50!, draw=cyan, line width=1pt,
    error bars/.cd, 
    x dir=both, 
    x explicit,
    error bar style={line width=.7pt},
    error mark options={mark size=2pt,line width=.7pt,rotate=90,cyan},] 
		coordinates{
(0.287, .4)+-(0.478, .4)
(1.216, .6)+-(0.517, .6)
(0.278, .8)+-(0.430, .8)
(0.416, 1)+-(0.418, 1)
}; \label{ExpProd}

\addplot+[fill=purple!60!, draw=purple, line width=1pt,
    error bars/.cd, 
    x dir=both, 
    x explicit,
    error bar style={line width=.7pt},
    error mark options={mark size=2pt,line width=.7pt,rotate=90, purple},] 
		coordinates{
(0.15, .2)+-(0.399,.2)
(-0.619, .4)+-(0.380, .4)
(-0.352, .6)+-(0.376, .6)
(-1.14, .8)+-(0.389, .8)
(-0.713, 1)+-(0.376, 1)
}; \label{VideoExpd}

\addplot+[
fill=pink!60!, draw=pink, line width=1pt,
    error bars/.cd, 
    x dir=both, 
    x explicit,
    error bar style={line width=.7pt},
    error mark options={mark size=2pt,line width=.7pt,rotate=90,pink},] 
		coordinates{
(0.094, .2)+-(0.322,.2)
(-0.576, .4)+-(0.485, .4) 
(0.25, .6)+-(0.477, .6) 
(-1.036, .8)+-(0.455, .8) 
(-0.791, 1)+-(0.429, 1)
}; \label{ExpRecd}

\draw[] (180,0.1) -- (180,1000);

\end{axis}
\end{tikzpicture}
\vspace{-0.7cm}\caption{Cohen's $d_p$ with pooled standard deviation \cite{GouletPelletier.2018} and 95\% confidence intervals \cite{Fritz.2012} %and $p$-values 
comparing affective responses to experimental, programming, and standard recitation tasks in this study, also with video analysis and recitation tasks from Ref.~\cite{Klein.2016}. Thereby, effect sizes are determined based on matched samples when comparing tasks within the presented study.  Dashed lines indicate small, medium, and strong effects. Effects sizes are comparable to those found in a meta-study on innovations in undergraduate science courses \cite{RuizPrimo.2011} for conceptually oriented tasks ($M=0.47$, $\mathrm{SD}=0.70$), using technology ($M=0.37$, $\mathrm{SD}=0.38$), or a combination of both ($M=0.41$, $\mathrm{SD}=0.61$).}
\label{fig:ComparisonReference}
\end{figure*}

To assess affective responses from the Exp-tasks in more detail, we consider effect sizes Cohen's $d_p$ with pool standard deviation \cite{GouletPelletier.2018} and 95\% confidence intervals \cite{Fritz.2012} of the post-hoc-comparisons between the Exp-, Pro-, and Rec-tasks. This comparison is supplemented by comparative data for weekly video analysis tasks and standard recitation tasks evaluated in a previous study\footnote{Reference~\cite{Klein.2016} investigated a group of 36 students who analyzed partly given, partly self-recorded videos on kinematics and dynamics and 40 control group students who completed comparable standard recitation tasks without video analysis. Despite some differences in data collection modes and rating scales, a comparison with our study is meaningful as data were collected in a comparable German first-semester physics course using similar items for \textit{curiosity}, \textit{interest}, \textit{reference to reality}, \textit{disciplinary authenticity}, and \textit{autonomy}.} in a similar setting using comparable instruments \cite{Klein.2016}. Table~\ref{Tab:overallcomparison} and effect sizes in Fig.~\ref{fig:ComparisonReference} reveal five key aspects for discussion.

1. Students in our study rated the Exp-tasks, in tendency, higher than the Pro-tasks for \textit{curiosity} ($d_p=0.42$, 95\%-CI $[0.00, 0.83]$), \textit{interest} ($d_p=0.28$, 95\%-CI $[-0.15, 0.71]$), and \textit{reference to reality} ($d_p=1.22$, 95\%-CI $[0.70, 1.73]$). This can be partially explained by open-text responses from the weekly surveys, where students described the Pro-tasks sometimes as difficult, rather complex, and (time-)demanding, with unclear instructions. Correspondingly, many expressed a desire for clearer guidance, more explanations, and a reduction in the scope and complexity of the tasks. This suggests that some students were probably challenged, possibly overwhelmed by the Pro-tasks, which may explain their lower affective responses to the Pro-tasks, as also reflected in the lower rating of the subscale \textit{curiosity/interest (short)}. In particular, the too-difficult task Pro~1 likely played a significant role in this outcome.

2. The Exp-tasks (and Pro-tasks) were perceived as lower than the Rec-tasks regarding \textit{curiosity} and \textit{interest}. This finding counters expectations about the use of smartphone experiments, as summarized in the introduction and state of research. Specifically, it differs from findings in Ref.~\cite{Kaps.2022}, where no significant difference in motivation and curiosity was found between students working on smartphone-based experimental tasks and those working on standard recitation tasks only. In their earlier work \cite{Kaps.2021}, students even reported increased interest in learning more about physics due to smartphone-based tasks. Several factors may explain the discrepancy between Exp- and Rec-tasks.

2.1. Open-text responses reveal that some students did not understand the added value or purpose of the tasks, found the Exp-tasks uninteresting or their instructions partly unclear. This may have influenced affective responses.

2.2. Although some students noted in open-text responses that they liked the simplicity and ease of certain tasks, this perception could suggest that these tasks were not sufficiently challenging or engaging. Notably, 24\% to 34\% of students rated three of the nine experimental tasks as too easy, indicating that these simpler tasks may not have adequately stimulated curiosity or sparked interest.

2.3. Low degrees of openness and novelty in some tasks (e.g., Exp~5, and Exp~7), serving more the confirmation than the exploration of physics content, may have similarly influenced affective responses. This is supported by open-text feedback for Exp~3, where students highly valued the creativity and freedom for own decisions about the setup and experimental design (63\% of according statements across all tasks).

2.4. When introducing new technologies, a possible effect of novelty on affective responses is often discussed \cite[e.g.,][]{Hochberg.2018,Kaps.2022}. Since many students already had prior experience with smartphone experiments (cf. Table~\ref{tab:priorexperience}) and the comparative surveys were conducted after completing several Exp-tasks, an initial novelty would have probably already been diminished by the time of measurement, likewise in Ref.~\cite{Kaps.2022}.

2.5. The survey Com~2 (Exp vs Rec) was conducted just days before the final exam, which focused solely on Rec-tasks. This timing may have amplified students' perceptions of aspects like \textit{authenticity}, but also \textit{perceived educational effectiveness}, or the tasks' \textit{linking to the lecture}.

2.6. Overall design of the learning environment may have influenced students' affective responses. For instance, the perceived relevance of the Exp- and Pro-tasks may have been reduced by their lower number compared to Rec-tasks and their exclusion from the final exam. The influence of the learning environment is also reflected in the perceived weaker \textit{linking to the lecture} of the Exp-tasks compared to the Rec-tasks.

3. Except for \textit{curiosity}, the affective responses to the standard recitation tasks were similar in both studies. These similarities support comparing the Exp-tasks in our study with the video analysis tasks in Ref.~\cite{Klein.2016} showing medium to strong negative effect sizes ranging from $d_p=-0.62$, 95\%-CI $[-1.00, -0.24]$, to $d_p=-1.14$, 95\%-CI $[-1.53, -0.75]$, for \textit{curiosity}, \textit{disciplinary authenticity}, and \textit{interest}. % These effect sizes are comparable to those between the Exp- and Rec-tasks since the video analysis tasks were rated rather similar to the standard recitation tasks in Ref.~\cite{Klein.2016}.
Thus, while new task formats can be successfully integrated into introductory physics courses, the potential of the Exp-tasks for affective responses has not been fully exploited in this study. The underlying reasons may be similar to those discussed for the differences between Exp- and Rec-tasks.

4. There are no significant differences in the scale \textit{autonomy} between the task types (as in the variable \textit{autonomy/creativity (short)}). This indicates a similar level of rather strong guidance, typical for shorter weekly exercises.

5. The \textit{disciplinary authenticity} of the Exp-tasks is, in tendency (but not significantly), and of the Pro-tasks significantly lower than for the Rec-tasks. This perception may arise from students' prior learning experiences also from school, where calculus- or algebra-based tasks dominate textbooks, exercises, and exams, where programming is rather uncommon for physics lessons, and where experiments are often demonstrated by teachers rather than conducted by students. Moreover, students might have expected to conduct experiments in a traditional lab setting rather than using smartphones at home. This is supported by open-text responses, where some students expressed dissatisfaction with the precision of smartphone experiments and a preference for better, more accurate equipment. Conversely, students perceived a high \textit{reference to reality} in the Exp-tasks, which is plausible as they were conducted at home or in everyday contexts, using students' personal smartphones and low-cost materials.

More generally, affective responses to smartphone-based experimental and programming tasks in introductory physics courses can also be limited by two further aspects: First, as also argued by Ref.~\cite{Organtini.2022} for the outcomes of their own smartphone-based experimental tasks, new, activating, and challenging teaching approaches that require high cognitive effort "may have a detrimental effect on students’ motivation, engagement, and ability to self-regulate their own learning" \cite[p.~19251]{Deslauriers.2019}, as it was shown for the comparison between passive and active instruction. Second, the first year of studying physics is often challenging for many students, leading to increased stress perceptions during almost the entire lecture and exam periods \cite{Lahme.2024c}; these circumstances may reduce students' openness to new task types and willingness to invest time and effort in tasks rather irrelevant to the final exam.

In conclusion, the discussion indicates that the smartphone-based Exp-tasks, mostly implemented for the first time, outperformed the new Pro-tasks and already achieved acceptable to good affective effects. Overall, their perception by students is positive (around 50\% to 80\% of the maximum value of the various scales), and comparable or only somewhat less positive than that for the well-established Rec-tasks. Note that the Rec-tasks have undergone repeated use and refinement over several years, suggesting that their improved perception among students may be due to the significant time and thought invested in their design. This interpretation aligns with the findings about video analysis tasks in Ref.~\cite{Klein.2016}, a special type of Exp-tasks, which also benefited from years of development and showed clear superiority compared to traditional tasks.

\subsection{Influence of predictors (RQ3)}

The analysis indicates that \textit{gender}, \textit{semester}, and \textit{study program} have minimal influence on students' perceptions of the three task types. Only higher-semester students have a significantly worse \textit{(P1) task impression} of the Pro-tasks than first-semester students, which could be due to two reasons: first, the Pro-tasks and a programming course were not mandatory for higher-semester students, whereas first-semester students had a mandatory programming course running in parallel, newly introduced to the physics study programs at the time of this study. Second, higher-semester students often retake the course, as they either did not write or failed the exam before, indicating that they may be weaker students overall. These factors combined could have contributed to higher-semester students finding the Pro-tasks more challenging than first-semester students.

The fact that all correlations between the parent variables \textit{(P1) task impression}, \textit{(P2) task value}, and \textit{(P3) affective impact} are significant (cf. Fig.~\ref{fig:Correlations}) suggests that the dependent variables in our study capture complementary and coherent aspects of task evaluation. The direction of these correlations indicates that students with a more favorable \textit{(P1) task impression} also tend to assign them higher \textit{(P2) task value} and more positive \textit{(P3) affective impact}, regardless of the task type.

The significant negative correlation between \textit{(P1) task impression} of the Exp-tasks and the \textit{high school graduation grade} might suggest that higher-achieving students may have more cognitive resources available to meet the task demands, resulting in a more favorable \textit{(P1) task impression}. The significant negative correlation between \textit{prior experience with physics experiments} and \textit{(P1) task impression} of the Pro-tasks could suggest two potential explanations: first, students with less experimental experience may find the Exp-tasks more challenging, leading them to prefer the Pro-tasks; second, students with more experimental experience might generally prefer hands-on experimental work, making them find the Pro-tasks less engaging or more difficult. The absence of a correlation between \textit{prior experience with physics/smartphone experiments} and \textit{(P1) task impression} of the Exp-tasks may be explained by the chosen task design, which provides detailed guidance, making the tasks accessible to students regardless of their prior experimental experience.

Overall, this part of the analysis shows that the considered predictors have rather little potential to explain differences in students' perceptions of learning with the different task types and the elicited affective responses. This is remarkable in that we observed rather higher standard deviations for the outcome variables examined. There are two possible explanations for this. First, our design as an exploratory field study with multiple surveys over the semester naturally led to a rather complex data structure with a high number of variables on the one hand and a relatively low number of complete responses in important surveys on the other. Thus, it may be that some of the predictors examined explain the differences in students' views, but that we were unable to demonstrate these relationships in our study. Second, there may be other predictors (e.g., openness to technology and learning innovations in general, or beliefs about learning physics) or covariates (e.g., related to task design, implementation, or overall course design) that are important in explaining the observed high standard deviations but have not been considered. Another interesting methodological approach (which would also require better dataset quality) would be the use of cluster analysis to analyze whether there are subgroups of students who tend to prefer a particular task type and why. The existence of such subgroups would be another possible explanation for the high variation in students' perceptions and could be the basis for internally differentiated learning opportunities.

\subsection{Limitations}

The following limitations of this study need to be acknowledged. First, there are limitations due to practical constraints. There were some inevitable delays between task completion and evaluation, so students sometimes started working on the next task before the previous one had been evaluated; this may have influenced task perceptions. Further, participation tended to decline across measuring points. This could be attributed to several factors, including survey fatigue, a typical drop-off in attendance at weekly recitation groups as the semester progresses, and the fact that students often fulfill the exam prerequisite before the semester ends, reducing their motivation to complete and evaluate all tasks. Moreover, as students worked in groups of three, not every group member may have responded to the survey individually. Additionally, not all variables could be measured for all task types, although they might have been interesting (e.g., \textit{perceived educational effectiveness} or \textit{autonomy} for the Pro-tasks). Finally, the Rec-tasks could be assessed only once at the end of the semester, meaning that students had to generalize across all Rec-tasks.

Second, there are limitations regarding instructional approach and setting entailed by the constraints of classroom research in a given teaching framework. In particular, this study is about a specific implementation of the Exp- and Pro-tasks, most of which were used for the first time. It was also carried out in a specific learning environment which is determined by other aspects such as the lectures or tutors' teaching skills. Thus, findings may have limited generalizability for different tasks, courses, student cohorts, or universities. Furthermore, we cannot distinguish to what extent the affective responses of the Exp-tasks are related to the use of smartphone experiments or Exp-tasks as weekly exercises in general (e.g., with other low-cost or standard lab equipment).

Third, this study addressed students' perceptions of learning with the tasks and affective responses using a specific and limited set of outcome variables relevant to the evaluation of smartphone-based experimental tasks and suitable for comparison with standard recitation and programming tasks. However, many other variables could be of interest, such as the influence on conceptual understanding \cite{Wieman.2015}, the acquisition of experimental skills \cite{Bauer.2023b}, scientific process skills \cite{Etkina.2006} like understanding measurement and handling measurement uncertainty \cite{Volkwyn.2008,Pollard.2020,Pillay.2008}, critical thinking skills \cite{Holmes.2015,Walsh.2019}, the development of beliefs about experimental physics and learning physics within lab courses \cite{Zwickl.2014,Wilcox.2016,Wilcox.2017}, or cultivating a greater physicist's habit, which combines several of these aspects \cite{Karelina.2007b}. Given the broad range of possible variables, some of which align with long lists of learning objectives for physics labs \cite[cf.][]{AmericanAssociationofPhysicsTeachers.1997,Welzel.1998b,Zwickl.2013,AmericanAssociationofPhysicsTeachers.2014}, not all variables can be addressed in a single study. Furthermore, it is necessary to consider whether all variables are suitable for comparing the three task types and whether and how they can actually be measured in the field over an entire semester, taking into account confoundings, e.g., caused by the accompanying lecture, the other task types, or the instruction in general. In particular, we consider evidence of conceptual or skill-related learning gains to be an interesting but difficult to measure outcome variable, as the methodological requirements hardly match the circumstances of real learning settings. For example, the acquisition of many competencies requires more time than a single or a few tasks, but their promotion over longer time is often difficult to implement in given course structures (e.g., with a rapid progression of learning content in the curriculum) and measured effects are difficult to attribute to instruction due to the many surrounding factors that may confound the results. Future research could address these and related challenges and show whether the smartphone-based experimental tasks can outperform standard recitation and programming with respect to other variables not considered in this study.

\section{Lessons learned and perspectives}

This manuscript contributes to the current state of research on smartphone-based experimental tasks in several ways that are relevant to both physics education researchers and instructors. First, our exploratory field study shows as a proof of concept that weekly smartphone-based experimental tasks can indeed be implemented in undergraduate physics education as weekly exercises over the course of a semester, which has been systematically reported by only a few authors so far, along with an evaluation only by Refs.~\cite{Kaps.2021,Kaps.2022}. In particular, students found our experimental tasks, most of which were implemented either for the first time or with significant changes compared to previous versions, sufficiently adequate. As logistical challenges such as providing experimental equipment to over a hundred student groups and integrating these tasks into assessments and exam prerequisites could be successfully overcome, smartphones and similar technologies do indeed facilitate student experimentation in undergraduate physics education beyond traditional laboratories. This feasibility may encourage other instructors to implement similar tasks in their own teaching, and also creates new opportunities for institutions where constraints such as limited equipment, space, or staff hinder traditional lab courses or other forms of student experimentation.

Second, the comparative analysis of affective responses showed that the smartphone-based experimental tasks were perceived as similar to or better than the programming tasks. They outperformed standard recitation tasks in terms of \textit{experience of competence}, but elicited less \textit{curiosity}, \textit{interest}, and \textit{perceived affective effectiveness}. These differences were only partially explained by student-related predictors, highlighting the role of task design and implementation.

Third, our study contributes methodologically to research on smartphone experiments by validating instruments for evaluating smartphone-based experimental tasks and by exemplifying a design for an exploratory evaluation study that can be implemented in a running course. In particular, the combination of short weekly surveys focused on individual tasks with more summative surveys comparing new and standard tasks has been shown to be an approach that provided versatile and meaningful data on student perceptions and task outcomes. The online surveys proved to be convenient for both instructors and students, requiring only 5~min for weekly surveys and 15~min for comparative surveys, i.e., for a total of about 90~min. High participation rates and comprehensive responses to open-text questions indicate that students are willing to provide feedback on new teaching approaches. Notably, some of the findings challenge common instructor expectations particularly about the affective potential of mobile devices in physics education, underscoring the importance of empirical evaluation for potential improvement. The scales analyzed in this study can assist other instructors in evaluating their own tasks and physics education researchers in supporting colleagues through external evaluations.

Overall, our study makes several important contributions to the development, implementation, systematic evaluation, and physics education research on the outcomes of technology-enhanced task formats in undergraduate physics education. %In particular, it provides a structured approach for integrating smartphone-based experimental tasks into regular teaching and shows that their systematic evaluation is both feasible and informative.
The comparative design of our evaluative study, going beyond a proof of concept, allows for a differentiated interpretation of students’ perceptions of smartphone-based experimental tasks in comparison to programming and standard recitation tasks. By evaluating and contrasting these three task types, we were able to assess not only the feasibility and perception of the experimental tasks but also their relative strengths and limitations in the broader instructional context. These findings provide valuable insights for instructors and curriculum developers who seek to diversify task formats in ways that support student engagement, autonomy, and learning motivation.

Concretely, our results provide a useful basis for future research that (i) bridges the gap between the large number of ideas for smartphone experiments and the comparatively small amount of research on their use in teaching, (ii) further investigates whether there are specific conditions (e.g., in task design or instructional environment) or other outcome variables (e.g., student achievement) in which smartphone-based experimental tasks might outperform traditional recitation tasks, and (iii) deepens our investigation of student predictors (e.g., gender, prior knowledge) on the perception of different task types. This can be achieved through studies with different designs and methods, including exploratory field studies like ours and more tightly controlled comparative studies. In particular, interviews and students' data and findings can also serve as valuable qualitative resources that provide deeper insights into students' task perceptions and their actual task processing and experimental processes. Furthermore, students' open-text feedback in our study can inform the development of a closed-ended questionnaire for a more systematic assessment of students' experiences. Future work should also explore the relationship between task design principles \cite[cf.][]{Lahme.2024b} and students' perceptions of learning with the tasks and affective responses. For instance, evaluation of smartphone-based experimental tasks as longer, open-ended undergraduate research projects \cite[cf.][]{Lahme.2024} may provide insights beyond the shorter weekly exercises examined in this study. 

All of these efforts will contribute to a new task culture in undergraduate physics courses that enhances traditional exercise sheets -- one of the most salient stressors among first-year physics students \cite[cf.][]{Lahme.2024c} -- with innovative elements such as student experimentation and use of technology, hopefully with a positive impact on students' cognition and affection.

%Although difficult to conduct, particularity such controlled comparative studies could extend our exploratory findings on the affective responses elicited by these tasks, and would allow for the examination of new variables, such as student performance, or a more reliable investigation of predictors such as gender and prior knowledge. 

\section{Ethical statement}

Data collection and storage were organized following the General Data Protection of the European Union and were coordinated with the data protection officer of the University of Göttingen. All participants were informed about the survey goals and the planned data collection and storage beforehand and consented to participate voluntarily and pseudonymously. For online surveys, in addition to an access time stamp, only personal data that participants were explicitly asked during the surveys were collected and stored. The entire data set of this study is openly available as supplementary material.

The authors D.~D., H.~H., C.~S., and S.~S. are among the developers of the \textit{phyphox} application used in this study to design and implement the experimental tasks.

ChatGPT, Grammarly, and DeepL were used to polish, condense, and lightly edit the pre-written text of this manuscript to improve its language and comprehensibility. The authors take full responsibility for the contents of the manuscript.

\section*{Acknowledgments}

This manuscript is part of the project Physik.SMART funded by the national foundation Stiftung Innovation in der Hochschullehre within the funding line Freiraum 2022. % We also acknowledge support by the Open Access Publication Funds/transformative agreements of the Göttingen University.
The authors thank Mosab Abumezied for his contributions and his thorough testing of student experiments. We would also like to thank Christian Effertz and Marina Hruska for testing the Bluetooth sensors at FH Aachen (University of Applied Sciences) as project partners. Further, we would like to thank our student assistant Laura Pflügl for her support in literature research and preparation of supplementary material.

\begin{footnotesize} D.~D.: conceptualization (supporting), resources (lead), writing – review \& editing (supporting); H.~H.: conceptualization (supporting), funding acquisition (equal), project administration, supervision (equal), writing – review \& editing (supporting); P.~K.: conceptualization (supporting), methodology (supporting), supervision (equal), writing – review \& editing (supporting); S.~Z.~L: conceptualization (lead), data curation, formal analysis, investigation (lead), methodology (lead), resources (supporting), validation, visualization, writing – original draft, writing – review \& editing (lead); A.~M.: conceptualization (supporting), methodology (supporting), supervision (equal), writing – review \& editing (supporting); S.~S.: conceptualization (supporting), funding acquisition (equal), investigation (supporting), resources (lead), writing – review \& editing (supporting); C.~S.: conceptualization (supporting), funding acquisition (equal), resources (lead).\end{footnotesize}

%\clearpage
\FloatBarrier
%\appendix*
\section*{Appendix}

\subsection*{Sample task instructions of Exp~5: \textit{Bouncing ball (7P)}}
\vspace{-12pt}
\noindent In your exercise group, you have been given a golf ball, a ping pong ball and a tape measure. Use a smartphone with the phyphox app and the experiment "(In)elastic collision" from the "Mechanics" category. When you start this (triangle in the top right-hand corner), the app can record the sound of a bouncing ball and calculate the height from which you dropped it. To do this, the app determines the time intervals between the sounds of the ball hitting the table. To do this, it may be necessary to increase the threshold under "Settings" and thus lower the sensitivity so that the app does not react to ambient noise, or vice versa to lower the threshold if the ball is not registered.
\begin{itemize}[align=left, topsep=0pt, itemsep=-12pt, partopsep=0pt, parsep=0pt,left=0pt,labelsep=3pt,]
\item[a)] Carry out the experiment with both balls and note down the time intervals measured and the heights calculated. (1P)\\
\item[b)] Derive a formula to determine the maximum height between two impacts using the time interval between them. Disregard the effects of friction. (1P)\\
\item[c)] The ball loses part of its energy with each impact (collision). Derive a formula with which you can calculate, based on two successive time intervals (corresponding to three impacts), the fraction of energy that is conserved during the second of the three impacts. From this, derive another formula that allows you to determine the original initial height $h_0$ before the first impact. Assume that the fraction of energy conserved by each impact is constant. (2P)\\
\item[d)] Verify your result by substituting the first time intervals from part a) and calculating the value for $h_0$ determined by phyphox. Using the other intervals from a), check whether the assumption that the conserved part of the energy is constant for each impact is correct. (1P)\\
\item[e)] Repeat the experiment but place a sheet of paper at the spot where the ball hits. Vary the number of sheets of paper (at least five different data sets per ball) and note how much energy the ball retains on the second impact. Plot your results as a function of the number of sheets of paper for both balls! (Make sure you drop the ball from the same height each time.) (2P)
\end{itemize}

%\onecolumngrid

\begin{table*}[htb]
\caption{Overview of the developed and implemented experimental and programming tasks. The last column indicates whether the tasks were used for the very first time (*), already in former cohorts but now with significant changes (**), or in the same way already before (***).}% Snippets of the tasks Exp~2, Exp~5, Exp~6, and Pro~3 are presented in Figs.~\ref{fig:Exp2}, \ref{fig:Exp5}, \ref{fig:Exp6}, and \ref{fig:Pro3}.}
\begin{ruledtabular}
\footnotesize
\begin{tabular}{p{.18\textwidth}p{.22\textwidth}p{.55\textwidth}p{.02\textwidth}}
Task (\& grading points)&Used technologies&Short description of the goals \& task instructions&\\\hline
\textbf{Experimental tasks (Exp)}&&&\\
\hangindent=.3cm Exp1 Drop cord with delayed time measuring (4P)&Acoustic stopwatch (phyphox)&\hangindent=.3cm Determination of the specific impact times and their proportion when a string with attached masses of equal distances falls onto a surface&*\\
\hangindent=.3cm Exp2 Sensor statistics of the smartphone (6P)&\hangindent=.3cm Accelerometer statistics (phyphox)&\hangindent=.3cm Mean, standard deviation and uncertainty of the mean of the noise data of the acceleration sensor for different measurement periods and smartphone angles&***\\
\hangindent=.3cm Exp3 Centrifugal acceleration on the smartphone (5P)&\hangindent=.3cm Centripetal acceleration (phyphox) \& Excel/Python for data analysis&\hangindent=.3cm Design \& conduction of an experiment to investigate the dependency of centrifugal acceleration \& angular velocity, visualization of the quadratic relationship, determination of the radius of rotation&***\\
\hangindent=.3cm Exp4 Friction on an inclined plane (7P)&\hangindent=.3cm Inclination (phyphox)&\hangindent=.3cm Determination of the coefficient of static friction between the smartphone \& three different surfaces of an inclined plane by measuring the angle when the smartphones starts to slide, followed by interpretation of second-hand data&***\\
Exp5 Bouncing ball (7P)&\hangindent=.3cm (In)elastic collision (phyphox)&\hangindent=.3cm Measurement of time intervals of a bouncing golf and table tennis ball, determination of the initial height from which the balls were dropped, analyzing the percentage of conserved energy per bounce, comparison of energy conservation for different piles of paper below the bouncing balls&**\\
\hangindent=.3cm Exp6 Moment of inertia on an inclined plane (6P)&\hangindent=.3cm Gyroscope sensor of an external sensor box (phyphox)&\hangindent=.3cm Determination of the moment of inertia of a wooden wheel on an inclined plane for various settings of attached screws based on the measured angular velocity over time&*\\
\hangindent=.3cm Exp7 Simple pendulum (4P) \cite[cf.][]{Gotze.2017}&\hangindent=.3cm Pendulum (phyphox)&\hangindent=.3cm Determination of the oscillation frequency of a smartphone attached to a string for different string lengths based on gyroscope data&**\\
\hangindent=.3cm Exp8 Pitot tube (4P) \cite[cf.][]{Dorsel.2022}&\hangindent=.3cm Experiment with pressure \& GPS sensors of an external sensor box (phyphox) \& Python for analysis&\hangindent=.3cm Comparing the velocity over time and its precision determined both by the pressure sensor in a moved Pitot tube (e.g., by running outside) and the velocity data derived from the GPS sensor&*\\
\hangindent=.3cm Exp9 Real damped oscillation (7P) \cite[cf.][]{Abumezied.2024}&\hangindent=.3cm Acceleration or light-distance sensor of an external sensor box (phyphox) \& Python for analysis&\hangindent=.3cm Measuring the oscillations of a spring pendulum with \& without attached damping cardboard, advanced data processing by using an offset for an oscillation around zero, by modeling the oscillation with an envelope function, and using a logarithmic scaling&*\\
\textbf{Programming tasks (Pro)}&&&\\
\hangindent=.3cm Pro1 Numerical path integral with Python (6P)&\hangindent=.3cm Prepared Python notebook&\hangindent=.3cm Set of shorter consecutive programming tasks to numerically determine a path integral for different functions including different paths in a gravitational field&*\\
\hangindent=.3cm Pro2 Relativistic journey (5P)&Prepared Python notebook&\hangindent=.3cm Implementing a board computer of a spacecraft to calculate travel times and distances at relativistic velocities converting between the inertial and solar frame of reference&*\\
\hangindent=.3cm Pro3 Simulated third Kepler's law (4P)&Prepared Python notebook&\hangindent=.3cm Set of shorter consecutive programming tasks to simulate planetary orbits and derive orbit parameters for the solar system&*\\% based on a function describing the current velocity \& position of an object in a simple throw \& gravitational acceleration of the Sun \\
\end{tabular}
\end{ruledtabular}
\label{tab:overviewtasks}
\end{table*}

\begin{table*}[htb]
\caption{Ordinal categorization of open-text responses about the $N=188$ students' \textit{prior expertise with physics experiments} in general and {smartphone experiments} in particular. When a participant listed multiple aspects belonging to different levels, the highest level was coded.}
\begin{ruledtabular}
\footnotesize
\begin{tabular}{p{.01\textwidth}p{.15\textwidth}p{.71\textwidth}r p{.07\textwidth}}
\multicolumn{2}{l}{Level of prior experience}&Description&$N$ students\\\hline
\multicolumn{3}{l}{\textbf{with physics experiments}}&\\
0&None&No, almost no, or mere experience with physics experiments&16\\
1&Only as demonstrations&No independent experimentation, only demonstrations in school or university&3\\
2&Basic from school&Independent experimentation in school lessons and school experiments in general&91\\
3&Advanced from school&Frequent independent experimentation in school, either in lessons or as part of school clubs, school projects, etc.&21\\
4&Outside of school&Independent experimentation outside of school, e.g. in physics tournaments or summer schools&28\\
5&Professional/Academic&Independent experimentation during previous professional training or studies at university&29\\
\multicolumn{3}{l}{\textbf{with smartphone experiments}}&\\
0&None&No prior experience with smartphones in the context of physics experiments&71\\
1&Just as an aiding tool&No use of built-in sensors or camera for physics experiments, but e.g. the calculator or stopwatch function&7\\
2&Basic&Use of built-in sensors and/or camera for physics experiments once or a few times&35\\
3&Advanced&Use of built-in sensors and/or camera for physics experiments multiple times, either in school or university&75
\end{tabular}
\end{ruledtabular}
\label{tab:priorexperience}
\end{table*}

\begin{table*}[htb]
\caption{Results of the exploratory factor analysis for the parent variables determined for RQ3 (cf. Sec.~\ref{MethodQuanAnalysis3}). Included variables and number of participants (cases), eigenvalues, and explained variance $\mathrm{Var}$ of each factor are given. % Only factors with eigenvalues $>1$ and an explained variance of at least 10\% were taken into account. 
For one-factor solutions, the minimal and maximal correlation with the factor across all variables is listed as minimal and maximal factor loads; otherwise, the according factor loads after varimax rotation are given. Further, z-standardized descriptive data (minimum $\mathrm{Min}$, maximum $\mathrm{Max}$, mean $M$, and standard deviation $\mathrm{SD}$) as well as Cronbach's $\alpha$, the maximum possible Cronbach's $\alpha_{max}$ if an item is deleted and the minimal item-total correlation $r_{i(t-i)}$ are given. Statistical parameters in \textit{italics} exceed or undermine thresholds recommended for a satisfying scale.}
\begin{ruledtabular}
\footnotesize
\begin{tabular}{p{.14\textwidth}p{.265\textwidth}p{.035\textwidth}p{.04\textwidth}p{.03\textwidth}p{.06\textwidth}p{.065\textwidth}p{.035\textwidth}p{.03\textwidth}p{.05\textwidth}p{.03\textwidth}p{.03\textwidth}p{.04\textwidth}p{.04\textwidth}}
\href{}{}(Sub-)Scale&Included variables&$N$ cases&Eigen-value&Expl. $\mathrm{Var}$&Min. factor load&Max. factor load&$\mathrm{Min}$&$\mathrm{Max}$&$M$&$\mathrm{SD}$&$\alpha$&$\alpha_{max}$&Min. $r_{i(t-i)}$\\\hline
(P1) Task impression&\hangindent=.3cm Overall task rating, goal clarity, feasibility at home, use of technologies&83&2.11&0.53&0.64&0.82&-2.00&2.00&0.02&0.66&0.69&0.69&0.38\\
Task value&\hangindent=.3cm Perceived affective \& cognitive effectiveness, autonomy, linking to the lecture&66&2.16&0.54&0.67&0.78&-2.00&1.00&0.02&0.74&0.72&0.69&0.43\\
$\mapsto$ (P2) Task value&\hangindent=.3cm \textit{Autonomy excluded for semantic reasons and to increase statistical quality}&66&1.86&0.62&0.76&0.82&-2.00&2.00&0.03&0.79&0.69&0.65&0.47\\
Affective impact&\hangindent=.3cm Curiosity, interest, disciplinary authenticity, reference to reality, experience of competence, curiosity/interest (short), autonomy/creativity (short)&113&4.18&0.60&0.26&0.92&-2.24&1.96&-0.011&0.75&0.87&\textit{0.91}&\textit{0.21}\\
$\mapsto$ (P3) Affective impact&\hangindent=.3cm \textit{Experience of competence excluded to increase statistical quality}&113&4.13&0.69&0.64&0.92&-2.74&2.02&-0.0074&0.83&0.91&0.92&0.53\\
\end{tabular}
\end{ruledtabular}
\label{tab:factoranalysisRG3}
\end{table*}

\begin{figure}[htb]
\flushleft 
\begin{tikzpicture}
\begin{axis}[width=.73\columnwidth, height=8.8cm, xbar=0pt,
  ymax=2.9,  
  xmin=1, xmax=10,
  xtick={1,2,3,4,5,6,7,8,9,10},
  ymin =0.5,
  xlabel={Mean \& standard deviation},
extra x ticks={1,10},
extra x tick labels={worst, best},
extra x tick style={grid=none, tick style={draw=none}, tick label style={xshift=-5pt, yshift=-8pt}},
  ytick = {.6,.8,1.,1.2,1.4,1.6,1.8,2.,2.2,2.4,2.6,2.8},
  yticklabel style={text width=.4\columnwidth,align=right, },
  yticklabels={Pro 3 (N=61), Pro 2 (N=88), Pro 1 (N=87), Exp 9 (N=37), Exp 8 (N=40), Exp 7 (N=68), Exp 6 (N=42), Exp 5 (N=81), Exp 4 (N=108), Exp 3 (N=92), Exp 2 (N=166), Exp 1 (N=188)},
   ytick pos=left,
xtick pos=left,
 legend columns=1, legend cell align = left,legend style = {draw = none},
 legend style = {at ={(-0.15,1)}, anchor = south west},
 bar width = 7pt,
  ]

\draw[ultra thick] (0,60) -- (1000,60);

\addplot+[purple!60!,area legend,
    draw=black, error bars/.cd, x dir=both, x explicit, error mark options={black,mark size=2pt,line width=.7pt,rotate=90
     },  error bar style={line width=.7pt}
      ] 
		coordinates{
(	5.77, .6)+-(2.27, .6)
(   7.07, .8)+-(2.07, .8)
(	5.56, 1)+-(2.39, 1)
(   6.41, 1.2)+-(2.24, 1.2)
(   6.63, 1.4)+-(1.97, 1.4)
(	6.24, 1.6)+-(1.78, 1.6)
(   6.52, 1.8)+-(1.92, 1.8)
(	6.19, 2)+-(2.16, 2)
(	6.64, 2.2)+-(1.88, 2.2)
(	6.74, 2.4)+-(2.47, 2.4)
(	6.41 ,2.6)+-(1.84, 2.6)
(	6.29, 2.8)+-(1.64, 2.8)
}; \label{overallLong}

\end{axis}
\end{tikzpicture}
\vspace{-0.7cm}\caption{Students' \textit{overall task rating} of each task.}% The dashed line indicates the scale middle.}
\label{fig:OverallImpressionLong}
\end{figure}
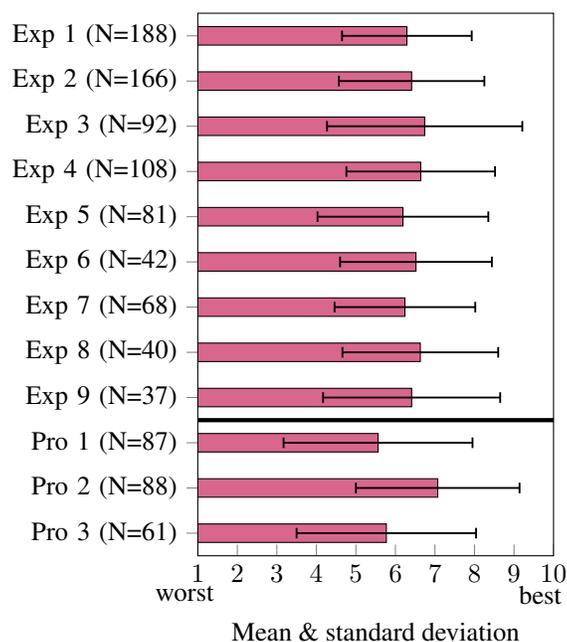

\begin{figure}[htb]
\flushleft 
\begin{tikzpicture}
\begin{axis}[width=.73\columnwidth, height=8.8cm, xbar=0pt,
  ymax=2.9,  
  xmin=0, xmax=310,
  xtick={0,60,120,180,240,300},
  ymin =0.5,
  xlabel={Time spent on task (in min)},
  ytick = {.6,.8,1.,1.2,1.4,1.6,1.8,2.,2.2,2.4,2.6,2.8},
  yticklabel style={text width=.4\columnwidth,align=right, },
  yticklabels={Pro 3 (N=53), Pro 2 (N=76), Pro 1 (N=76), Exp 9 (N=28), Exp 8 (N=33), Exp 7 (N=60), Exp 6 (N=38), Exp 5 (N=77), Exp 4 (N=102), Exp 3 (N=79), Exp 2 (N=156), Exp 1 (N=175)},
   ytick pos=left,
xtick pos=left,
 legend columns=1, legend cell align = left,legend style = {draw = none},
 legend style = {at ={(-0.15,1)}, anchor = south west},
 boxplot/draw direction=x,
 boxplot={
        box extend=0.1pt
    }
  ]

\draw[ultra thick] (0,60) -- (1000,60);

\addplot[
 boxplot prepared={
    lower whisker=10,
    lower quartile=30,
    median=45,
    upper quartile=75,
    upper whisker=300,
    },
     boxplot/draw position=0.6,
    ] coordinates {};%Pro3

\addplot[
 boxplot prepared={
    lower whisker=5,
    lower quartile=15,
    median=30,
    upper quartile=46.25,
    upper whisker=240,
    },
    boxplot/draw position=0.8,
    ] coordinates {};%Pro2

\addplot[
 boxplot prepared={
    lower whisker=20,
    lower quartile=60,
    median=90,
    upper quartile=120,
    upper whisker=300,
    },
    boxplot/draw position=1,
    ] coordinates {};%Pro1

\addplot[
 boxplot prepared={
    lower whisker=15,
    lower quartile=30,
    median=50,
    upper quartile=97.5,
    upper whisker=200,
    },
    boxplot/draw position=1.2,
    ] coordinates {};%Exp9

    \addplot[
 boxplot prepared={
    lower whisker=10,
    lower quartile=30,
    median=45,
    upper quartile=60,
    upper whisker=220,
    },
    boxplot/draw position=1.4,
    ] coordinates {};%Exp8

    \addplot[
 boxplot prepared={
    lower whisker=7,
    lower quartile=30,
    median=40,
    upper quartile=60,
    upper whisker=180,
    },
    boxplot/draw position=1.6,
    ] coordinates {};%Exp7

    \addplot[
 boxplot prepared={
    lower whisker=7,
    lower quartile=32.5,
    median=60,
    upper quartile=90,
    upper whisker=200,
    },
    boxplot/draw position=1.8,
    ] coordinates {};%Exp6

    \addplot[
 boxplot prepared={
    lower whisker=5,
    lower quartile=30,
    median=40,
    upper quartile=60,
    upper whisker=240,
    },
    boxplot/draw position=2,
    ] coordinates {};%Exp5

    \addplot[
 boxplot prepared={
    lower whisker=5,
    lower quartile=15,
    median=30,
    upper quartile=44.675,
    upper whisker=120,
    },
    boxplot/draw position=2.2,
    ] coordinates {};%Exp4

        \addplot[
 boxplot prepared={
    lower whisker=5,
    lower quartile=18.75,
    median=30,
    upper quartile=41.25,
    upper whisker=120,
    },
    boxplot/draw position=2.4,
    ] coordinates {};%Exp3
    
    \addplot[
 boxplot prepared={
    lower whisker=5,
    lower quartile=15,
    median=20,
    upper quartile=30,
    upper whisker=180,
    },
    boxplot/draw position=2.6,
    ] coordinates {};%Exp2
    
    \addplot[
 boxplot prepared={
    lower whisker=5,
    lower quartile=15,
    median=30,
    upper quartile=50,
    upper whisker=150,
    },
    boxplot/draw position=2.8,
    ] coordinates {};%Exp1
    
\begin{comment}\addplot+[purple!60!,area legend,
    draw=black, error bars/.cd, x dir=both, x explicit, error mark options={black,mark size=2pt,line width=.7pt,rotate=90
     },  error bar style={line width=.7pt}
      ] 
		coordinates{
(	68.36, .6)+-(79.10, .6)
(   45.24, .8)+-(80.46, .8)
(	104.35, 1)+-(78.23, 1)
(   62.57, 1.2)+-(72.63, 1.2)
(   54.19, 1.4)+-(50.79, 1.4)
(	44.10, 1.6)+-(33.01, 1.6)
(   65.87, 1.8)+-(39.14, 1.8)
(	53.03, 2)+-(47.27, 2)
(	32.25, 2.2)+-(23.40, 2.2)
(	39.89, 2.4)+-(67.97, 2.4)
(	23.54 ,2.6)+-(21.48, 2.6)
(	33.79, 2.8)+-(26.01, 2.8)
}; \label{TimeLong}
\end{comment}
\end{axis}
\end{tikzpicture}
\vspace{-0.7cm}\caption{Students' self-reported time spent on the different tasks presented as boxplots. The data was winsorized beforehand, i.e. responses below 5~min or above 300~min are excluded.}%The dashed lines indicate time durations of 30~min, 60~min, and 90~min.}
\label{fig:TimeLong}
\end{figure}
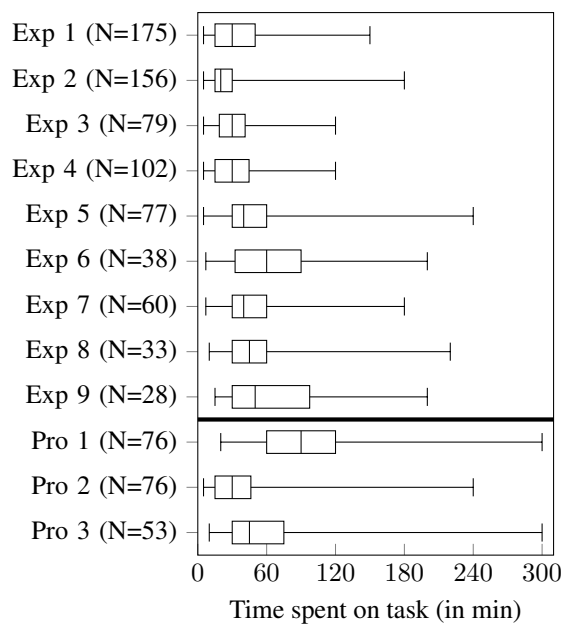

\begin{figure}[htb]
\flushleft 
\begin{tikzpicture}
\begin{axis}[width=.73\columnwidth, height=9.5cm, xbar=0pt,
  ymax=2.9,  
  xmin=1, xmax=5,
  xtick={1,2,3,4,5},
  ymin =0.5,
  xlabel={Mean \& standard deviation},
extra x ticks={1,5},
extra x tick labels={strongly disagree, strongly agree},
extra x tick style={grid=none, tick style={draw=none}, tick label style={xshift=-25pt, yshift=-8pt}},
  ytick = {.6,.8,1.,1.2,1.4,1.6,1.8,2.,2.2,2.4,2.6,2.8},
  yticklabel style={text width=.4\columnwidth,align=right, },
  yticklabels={Pro 3 (N=57), Pro 2 (N=84), Pro 1 (N=80), Exp 9 (N=39), Exp 8 (N=37), Exp 7 (N=63), Exp 6 (N=38), Exp 5 (N=79), Exp 4 (N=103), Exp 3 (N=80), Exp 2 (N=162), Exp 1 (N=181)},
   ytick pos=left,
xtick pos=left,
 legend columns=1, legend cell align = left,legend style = {draw = none},
 legend style = {at ={(-0.19,1)}, anchor = south west},
 bar width = 5pt,
  ]
\addlegendimage{empty legend}
\addlegendentry{\scalebox{1}[1]{\ref{ClarityLong}} Goal clarity}
\addlegendimage{empty legend}
\addlegendentry{\scalebox{1}[1]{\ref{FeasibilityLong}} Feasibility at home}

\draw[ultra thick] (0,60) -- (1000,60);

\addplot+[purple!30!,area legend,
    draw=black, error bars/.cd, x dir=both, x explicit, error mark options={black,mark size=2pt,line width=.7pt,rotate=90
     },  error bar style={line width=.7pt}
      ] 
		coordinates{
(	3.69, .6)+-(0.97, .6)
(   4.09, .8)+-(0.92, .8)
(	3.73, 1)+-(0.89, 1)
(   4.02, 1.2)+-(0.88, 1.2)
(   3.82, 1.4)+-(0.78, 1.4)
(	3.91, 1.6)+-(0.81, 1.6)
(   4.11, 1.8)+-(0.97, 1.8)
(	4.07, 2)+-(0.75, 2)
(	4.15, 2.2)+-(0.78, 2.2)
(	3.74, 2.4)+-(0.97, 2.4)
(	4.36 ,2.6)+-(0.74, 2.6)
(	4.04, 2.8)+-(0.72, 2.8)
}; \label{FeasibilityLong}

\addplot+[purple!60!,area legend,
    draw=black, error bars/.cd, x dir=both, x explicit, error mark options={black,mark size=2pt,line width=.7pt,rotate=90
     },  error bar style={line width=.7pt}
      ] 
		coordinates{
(	3.71, .6)+-(0.81, .6)
(   3.99, .8)+-(0.72, .8)
(	3.52, 1)+-(0.69, 1)
(   4.00, 1.2)+-(0.76, 1.2)
(   4.07, 1.4)+-(0.65, 1.4)
(	4.06, 1.6)+-(0.57, 1.6)
(   3.91, 1.8)+-(0.72, 1.8)
(	4.00, 2)+-(0.61, 2)
(	4.15, 2.2)+-(0.61, 2.2)
(	4.09, 2.4)+-(0.60, 2.4)
(	4.02 ,2.6)+-(0.67, 2.6)
(	3.92, 2.8)+-(0.65, 2.8)
}; \label{ClarityLong}

\end{axis}
\end{tikzpicture}
\vspace{-0.7cm}\caption{\textit{Goal clarity} and \textit{feasibility at home} perceived by the students comparing all twelve tasks.}% The dashed line indicates an agreement.}
\label{fig:AdequacyscaleLong}
\end{figure}
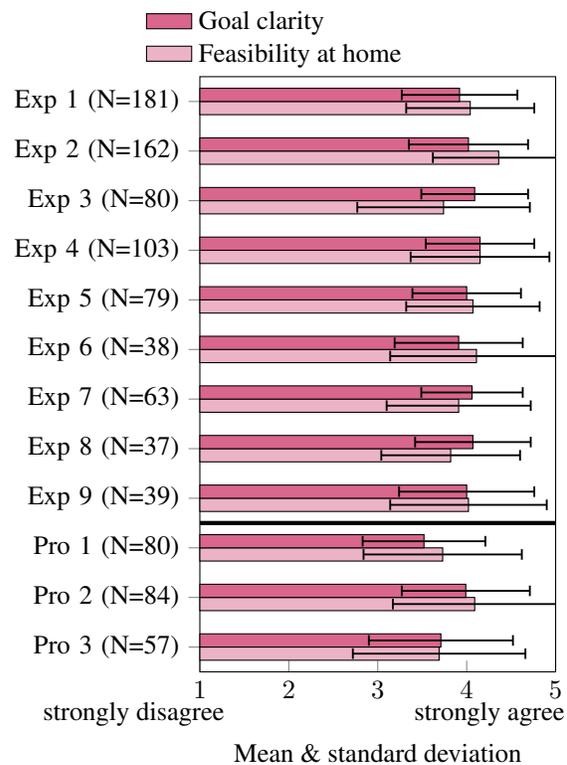

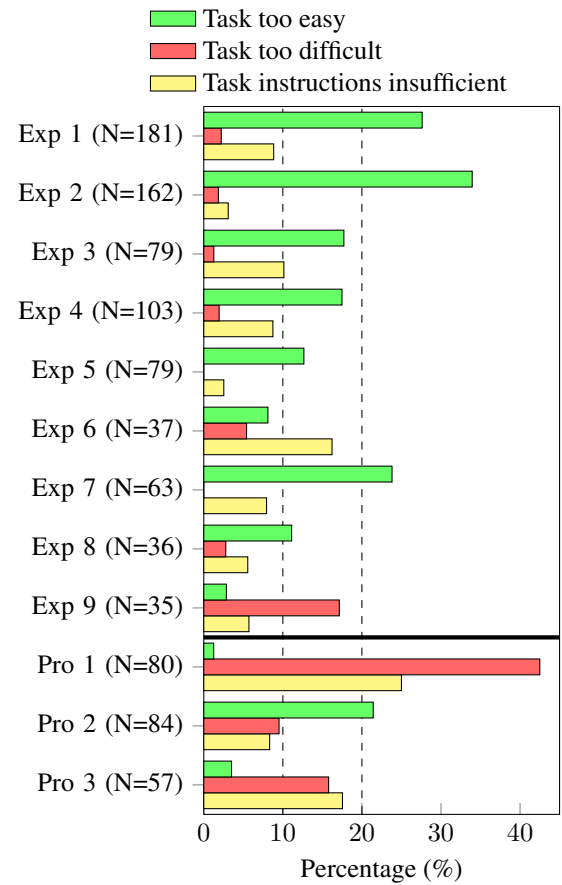
\begin{figure}[htb]
\flushleft 
\begin{tikzpicture}
\begin{axis}[width=.73\columnwidth, height=11cm, xbar=0pt,
  ymax=2.9,  
  xmin=0, xmax=45,
  xtick={0,10,20,30,40},
  ymin =0.5,
  xlabel={Percentage (\%)},
  ytick = {.6,.8,1.,1.2,1.4,1.6,1.8,2.,2.2,2.4,2.6,2.8},
  yticklabel style={text width=.4\columnwidth,align=right, },
  yticklabels={Pro 3 (N=57), Pro 2 (N=84), Pro 1 (N=80), Exp 9 (N=35), Exp 8 (N=36), Exp 7 (N=63), Exp 6 (N=37), Exp 5 (N=79), Exp 4 (N=103), Exp 3 (N=79), Exp 2 (N=162), Exp 1 (N=181)},
   ytick pos=left,
xtick pos=left,
 legend columns=1, legend cell align = left,legend style = {draw = none},
 legend style = {at ={(-0.19,1)}, anchor = south west},
 bar width = 6pt,
  ]
\addlegendimage{empty legend}
\addlegendentry{\scalebox{1}[1]{\ref{EasyLong}} Task too easy}
\addlegendimage{empty legend}
\addlegendentry{\scalebox{1}[1]{\ref{DifficultLong}} Task too difficult}
\addlegendimage{empty legend}
\addlegendentry{\scalebox{1}[1]{\ref{InsufficientLong}} Task instructions insufficient}

\draw[dashed] (100,0.1) -- (100,400);
\draw[dashed] (200,0.1) -- (200,400);

\draw[ultra thick] (0,60) -- (600,60);

\addplot+[yellow!60!,area legend,
    draw=black, error bars/.cd, x dir=both, x explicit, error mark options={black,mark size=2pt,line width=.7pt,rotate=90
     },  error bar style={line width=.7pt}
      ] 
		coordinates{
(	17.54, .6)
(   8.33, .8)
(	25.00, 1)
(   5.71, 1.2)
(   5.56, 1.4)
(	7.94, 1.6)
(   16.22, 1.8)
(	2.53, 2)
(	8.74, 2.2)
(	10.13, 2.4)
(	3.09,2.6)
(	8.84, 2.8)
}; \label{InsufficientLong}

\addplot+[red!60!,area legend,
    draw=black, error bars/.cd, x dir=both, x explicit, error mark options={black,mark size=2pt,line width=.7pt,rotate=90
     },  error bar style={line width=.7pt}
      ] 
		coordinates{
(	15.79, .6)
(   9.52, .8)
(   42.50, 1)
(   17.14, 1.2)
(   2.78, 1.4)
(	0.00, 1.6)
(   5.41, 1.8)
(	0.00, 2)
(	1.94, 2.2)
(	1.27, 2.4)
(	1.85 ,2.6)
(	2.21, 2.8)
}; \label{DifficultLong}

\addplot+[green!60!,area legend,
    draw=black, error bars/.cd, x dir=both, x explicit, error mark options={black,mark size=2pt,line width=.7pt,rotate=90
     },  error bar style={line width=.7pt}
      ] 
		coordinates{
(	3.51, .6)
(   21.43, .8)
(	1.25, 1)
(   2.86, 1.2)
(   11.11, 1.4)
(	23.81, 1.6)
(   8.11, 1.8)
(	12.66, 2)
(	17.48, 2.2)
(	17.72, 2.4)
(	33.95 ,2.6)
(	27.62, 2.8)
}; \label{EasyLong}

\end{axis}
\end{tikzpicture}
\vspace{-0.7cm}\caption{Percentage of students perceiving the according task as too easy or difficult and the task instructions as insufficient.}
\label{fig:difficultyInstructionslong}
\end{figure}

\begin{table*}[htb]
\caption{Overview of Bonferroni-corrected Friedman-tests with test statistic $\chi^2$ comparing the three task types regarding the affective responses (RQ2). In case of significant differences, Bonferroni-corrected post-hoc tests are conducted, whereby the test statistic $z$ represents the standardized difference between mean ranks of the two compared task formats.}%Bonferroni 4x, 3x
\begin{ruledtabular}
\footnotesize
\begin{tabular}{p{.18\textwidth}p{.05\textwidth}p{.05\textwidth} p{.05\textwidth} p{.02\textwidth}p{.2\textwidth}p{.2\textwidth}p{.2\textwidth}}
Variable&$\chi^2(2)$&$p$&$p_B$&$N$&Post-hoc Exp vs Pro&Post-hoc Exp vs Rec&Post-hoc Pro vs Rec\\\hline
Curiosity&34.08&<0.001&<0.001&45&$z=2.32$,\newline$p=0.020$, $p_B=0.061$&$z=-3.43$,\newline$p<0.001$, $p_B=0.002$&$z=-5.75$,\newline$p<0.001$, $p_B<0.001$\\
Interest&28.8&<0.001&<0.001&42&$z=1.26$,\newline$p=0.21$, $p_B=0.63$&$z=-3.87$,\newline$p<0.001$, $p_B<0.001$&$z=-5.13$,\newline$p<0.001$, $p_B<0.001$\\
Reference to reality&24.1&<0.001&<0.001&34&$z=4.49$,\newline$p<0.001$, $p_B<0.001$&$z=0.61$,\newline$p=0.54$, $p_B=1.00$&$z=-3.88$,\newline$p<0.001$, $p_B<0.001$\\
Disciplinary authenticity&11.3&0.004&0.014&34&$z=0.97$,\newline$p=0.33$, $p_B=1.00$&$z=-2.24$,\newline$p=0.025$, $p_B=0.075$&$z=-3.21$,\newline$p=0.001$, $p_B=0.004$\\
Experience of competence&15.0&<0.001&0.002&31&$z=3.49$,\newline$p<0.001$,$p_B=0.001$&$z=2.99$,\newline$p=0.003$, $p_B=0.009$&$z=-0.51$,\newline$p=0.61$, $p_B=1.00$\\
Curiosity/Interest (short)&8.5&0.014&0.042&31&$z=1.14$,\newline$p=0.25$, $p_B=0.76$&$z=-1.52$,\newline$p=0.13$, $p_B=0.38$&$z=-2.67$,\newline$p=0.008$, $p_B=0.023$\\
Autonomy/Creativity (short)&1.5&0.48&1.00&31&&&\\%&$z=-0.27$, $p=$, $p_B=0.84$&$z=-0.065$, $p_B=1.00$&$z=-0.21$, $p_B=1.00$\\
\end{tabular}
\end{ruledtabular}
\label{Tab:FriedmanAffec}
\end{table*}

%\FloatBarrier
%\clearpage
%\twocolumngrid
\bibliography{bibtex1}

\end{document}